\documentclass[acmsmall]{acmart}
\AtBeginDocument{%
  }

\usepackage{xcolor}

\received{13 May 2025}
\received[revised]{13 January 2026}
\received[accepted]{9 April 2026}

\begin{document}

\title{“The Only Thing Certain About This is Uncertainty”: Exploring Informal Care Coordination Practices Among Older Adults with Mild Cognitive Impairment}

\author{Josey M. Benandi}
\authornote{Both authors contributed equally to this research.}
\affiliation{%
  \institution{Georgia Institute of Technology}
  \city{Atlanta}
  \state{Georgia}
  \country{USA}
}

\author{Niharika Mathur}
\email{nmathur35@gatech.edu}
\authornotemark[1]
\affiliation{%
  \institution{Georgia Institute of Technology}
  \city{Atlanta}
  \state{Georgia}
  \country{USA}
}

\author{Sangha Park}
\affiliation{%
  \institution{Georgia Institute of Technology}
  \city{Atlanta}
  \state{Georgia}
  \country{USA}
}

\author{Tracy L. Mitzner}
\affiliation{%
  \institution{Georgia Institute of Technology}
  \city{Atlanta}
  \state{Georgia}
  \country{USA}
}

\author{Elizabeth D. Mynatt}
\affiliation{%
  \institution{Northeastern University}
  \city{Boston}
  \state{Massachusetts}
  \country{USA}
}

\author{Agata Rozga}
\affiliation{%
  \institution{Georgia Institute of Technology}
  \city{Atlanta}
  \state{Georgia}
  \country{USA}
}

\renewcommand{\shorttitle}{Informal Care Coordination Practices Among Older Adults with MCI}
\renewcommand{\shortauthors}{Josey M. Benandi et al.}

\begin{abstract}
Older adults aging in place often have informal support systems to help them \textcolor{black}{maintain} independence and quality of life. As they age, many older adults deal with the onset of Mild Cognitive Impairment (MCI), which introduces a new set of functional and cognitive changes \textcolor{black}{that affect their ability to manage daily routines}. \textcolor{black}{The approach to arranging and coordinating support for everyday activities for older adults with MCI varies across informal care networks, but typically involves a primary care partner and a network of} family, friends and others. In this paper, we present a thematic \textcolor{black}{analysis} of in-depth interviews with older adults with MCI \textcolor{black}{and their primary care partners to gain a holistic picture of their day-to-day lived experience}. \textcolor{black}{Our} \textcolor{black}{analysis} \textcolor{black}{uses a multi-dimensional lens of} people (“\textit{who}”), activities (“\textit{what}”), and tools (“\textit{how}”) \textcolor{black}{to reveal insights about the nature of informal care coordination in MCI. Our results characterize informal care for MCI as a set of complex orchestration tasks by a primary care partner that support the practical, cognitive and emotional needs of the diagnosed individual and mediate the involvement of the broader care network. We uncover that coordination is not solely a matter of logistical organization, but also a deeply relational process shaped by negotiation with technological tools and evolving roles. Through this work, we reframe care coordination for MCI as a distinct and underexplored design space, one that demands systems capable of scaffolding autonomy, adapting to shifting capacities, \textcolor{black}{responding} to socio\textcolor{black}{-}emotional needs, and fostering collaborative caregiving.} 
\end{abstract}

\setcopyright{cc}
\setcctype{by}
\acmJournal{PACMHCI}
\acmYear{2026} \acmVolume{10} \acmNumber{6} \acmArticle{CSCW165}
\acmMonth{10} \acmDOI{10.1145/3817013}



\maketitle

\section{Introduction}

\textcolor{black}{The work presented herein seeks to unpack the nature of care coordination for individuals diagnosed with Mild Cognitive Impairment (MCI) in order to identify new opportunities for designing technological supports for older adults with MCI and their informal care networks.} MCI is a clinical diagnosis characterized by the onset of cognitive decline, including challenges with memory, executive function, and language difficulties that exceed what is typically observed in individuals of a similar age undergoing the normal process of aging \cite{petersen1997aging, petersen2014mild}. The incidence of MCI continues to rise; a recent meta-analysis suggests that worldwide, more than 19\% of people have MCI with a significant rise in prevalence after 2019 \cite{song2023evidence}. \textcolor{black}{Critically, individuals with MCI are at elevated risk for progressing into dementia, with conversion rates of 10-15\% per year compared with 1-2\% among healthy older adults \cite{winblad2016defeating, wang2022management}}. MCI thus presents a critical inflection point in the aging trajectory, underscoring the importance of timely interventions \textcolor{black}{to maximize autonomy and independence}. 

This understanding motivates our work, which aims to identify opportunities for designing technological supports for older adults with MCI and their informal caregivers\textcolor{black}{\footnote{\textcolor{black}{We will refer to these informal caregivers as “care partners” or “CPs” throughout this paper in recognition of the collaborative partnership between the diagnosed individual and the informal care provider, and in line with recent trends in the MCI literature \cite{mathur2022collaborative, zubatiy2021empowering, zubatiy2023don}.}}} \textcolor{black}{- typically spouses but often also adult children and other family members - \textcolor{black}{as they} navigate these cognitive changes and their impact on daily routines and needs for support.}
While the experiences of people with MCI vary widely based on the progression and level of impact, \textcolor{black}{these individuals} often retain the ability to \textcolor{black}{independently} engage in \textcolor{black}{basic activities of daily living such as dressing, grooming, and feeding. Instead, they exhibit subtle decline in their performance of more cognitively challenging day-to-day tasks, such as managing their medications and finances, using the telephone, and using public or private transportation} \cite{petersen1997aging}. Prior research into the lived experiences of those diagnosed with MCI also shows that informal care partners, such as spouses or other family members, play a key role in supporting them as they navigate \textcolor{black}{these challenges, scaffolding complex tasks to maximize their loved ones’ independence} \cite{mank2023determinants}. Furthermore, research shows that more than 60\% of individuals \textcolor{black}{with MCI will} advance to dementia within six years of being diagnosed \cite{busse2006progression}. Given the substantial impact of dementia on reduced autonomy and increased reliance on formal caregivers, this period between healthy aging and dementia onset presents a critical opportunity to introduce tools or technological interventions that promote independence \cite{anderson2019state, liang2019optimal}. Older adults with MCI may benefit from cognitive compensatory strategies - assistive support activities executed by others that allow them to engage in daily activities requiring high cognitive effort - that can play a vital role in preserving functional abilities. These strategies often span both human and technological forms of support, aiming not just to reduce cognitive load but to maintain engagement and agency \cite{mathur2022collaborative, sherman2017efficacy}. As such, socio-technical tools that align with the lived realities of older adults and their CPs offer promise for enhancing quality of life and supporting aging-in-place.

Our work leverages \textcolor{black}{in-depth, semi-structured interviews and qualitative analysis methods} to investigate \textcolor{black}{the nature of informal care coordination in MCI; specifically,} how older adults with MCI and their informal care \textcolor{black}{partners navigate and coordinate daily routines and any compensatory practices they rely on. Care coordination is a broad term encompassing both formal care, in which healthcare professionals collaborate to manage a patient's healthcare journey, as well as informal care, in which unpaid family or friends organize and manage care for a loved one, often alongside formal healthcare. While prior work has focused extensively on both formal and informal care coordination for individuals with dementia, less is known about care coordination for individuals with MCI who, as noted above, continue to largely rely on informal caregivers for support.}

\textcolor{black}{In this work, }we address \textcolor{black}{these gaps through} \textcolor{black}{two} primary research questions: 
\textcolor{black}{
\begin{itemize}
    \item \textcolor{black}{\textbf{RQ1:}} How is the everyday care coordination for older adults with MCI structured, and what are the roles and strategies of informal care partners in enabling this coordination?
    \item \textbf{RQ2:} What tools and systems \textcolor{black}{do} older adults with MCI and their CPs \textcolor{black}{use} to manage and coordinate daily activities, and \textcolor{black}{what are the shortcomings of} these tools \textcolor{black}{in} meeting their coordination needs? 
\end{itemize}
}
\subsection{Contributions}

\textcolor{black}{A key contribution of our work is a set of insights into the nature of informal care coordination for older adults with MCI framed along three dimensions:} (1) what activities are coordinated, (2) the tools used to coordinate those activities, and (3) the people involved in coordination. We characterize the space of daily activities and routines that require coordination, the unique ways both analog and digital tools are used to support this coordination, and identify the key stakeholders who form the informal care networks of older adults with MCI. 

\textcolor{black}{We introduce the notion of “orchestration” as a key facet of how informal care coordination for MCI is manifested. This term is intended to capture our key finding that for individuals with MCI, a single primary care partner, typically the spouse, takes on the main responsibility for providing and/or arranging the necessary scaffolds, supports, and opportunities needed to maximize the independence of their loved one with MCI, including facilitating the participation of other \textcolor{black}{formal} and informal caregivers in care activities. Notably, this orchestration occurs not only in the context of practical support for daily routines, but also in support of the emotional needs of the loved one with MCI, such as their desire for autonomy and independence.} \textcolor{black}{These insights have} the potential to serve as foundations for future work on technology-mediated care coordination for individuals with MCI. \textcolor{black}{Our characterization of care coordination for MCI as an orchestration activity by a primary CP that \textcolor{black}{exhibits} acute emotional awareness presents the  CSCW community with new opportunities for developing socio-technical systems that can more robustly support the needs of MCI specifically and older adults more broadly. Framing coordination technology around socio-emotional considerations for a diagnosis that is progressive and that relies on non-medical interventions and informal care \textcolor{black}{brings to light opportunities for solutions that} will support dyads and their networks in ways that current technologies simply cannot. \textcolor{black}{For our purposes, socio-emotional considerations are defined as the continued awareness of and action taken for the sake of one’s mental wellbeing and enrichment, so as to prioritize personal empowerment in the provision of supports.}}

\textcolor{black}{Finally,} our work makes the following \textcolor{black}{additional} contributions to the HCI and CSCW research communities:

\begin{itemize}
    \item \textbf{Empirical contribution:} We provide rich, real-world accounts of the lived experiences of older adults with Mild Cognitive Impairment (MCI) and their informal CPs, touching upon their needs for \textcolor{black}{orchestrated} support and the uncertainty that comes with such a diagnosis.
    \item \textbf{Design implications}: Based on these experiential accounts, we provide insights to inform the design of future socio-technical systems to support collaborative care coordination in home settings, enabling successful practices for aging in place for older adults with MCI.
\end{itemize}

\section{Background}

\subsection{\textcolor{black}{Informal Care Coordination}}

\textcolor{black}{Care coordination is often treated as a broad umbrella within CSCW and HCI, encompassing a wide range of practices across chronic health conditions such as diabetes, mental health conditions, and physical disability, focusing primarily on coordination of everyday tasks \cite{foong2020you, schurgin2021isolation}. In such contexts, coordination work frequently centers on managing medically oriented tasks, including scheduling clinical appointments \cite{zubatiy2021empowering}, monitoring symptoms \cite{ullgren2018family, dawber2019comparison}, adhering to treatment regimens \cite{knodel2010role, zhou2025adhera}, and coordinating with healthcare professionals \cite{miller2016partners}. Even when caregiving occurs in informal or home-based settings, the locus of coordination is often shaped by healthcare systems and clinical requirements, such as medication management, insulin monitoring or information exchange with providers. As a result, existing care coordination frameworks are deeply informed by assumptions of ongoing medical engagement, formal care infrastructures, and clearly defined health-related tasks. While these accounts provide valuable insight into the cognitive and emotional labor of caregiving, they do not fully account for coordination contexts in which medical care is not central, but where everyday functioning, anticipation, and relational negotiation become the primary sites of care work (such as MCI, discussed more in 2.2).}

\textcolor{black}{Previous works have also addressed and explored the emotional impact that providing care has on caregivers \cite{bhat2023we, nikkhah2024family, chen2013caring}, highlighting the emotional support caregivers provide in helping the cared-for individuals navigate the psychological and emotional effects of living with a chronic condition \cite{miller2016partners}. These studies characterize informal care as a complex cognitive and emotional activity that demands an acute awareness and attunement to the needs of the cared-for individual while exacting a personal toll on informal caregivers witness\textcolor{black}{ing} a loved \textcolor{black}{one} dealing with illness.} \textcolor{black}{This caregiving experience is \textcolor{black}{often noted} by individuals providing care for loved ones diagnosed with dementia, with the added complexity that caregivers in this context are also navigating challenges that come with managing cognitive decline \cite{ducharme2011learning, zarit2008behavioral}. In these instances, coordinating care takes on an added dimension of not only managing medical care and supporting daily activities \textcolor{black}{such as} grooming, dressing, feeding, etc., but also addressing the memory challenges and psychological ramifications (i.e., personality changes and increased emotional volatility) that accompany the diagnosis. This multi-fold approach is unique to caregiving in the context of cognitive decline, as interventional support must prioritize the physical, cognitive, and socio-emotional dimensions  of maintaining independence and quality of life \cite{kim2025evaluation}.}

\textcolor{black}{Beyond documented experiences of individual caregivers, prior work has also explored the} interpersonal \textcolor{black}{and collaborative} nature of informal caregiving, emphasizing distributed coordination among people, artifacts and information flows \cite{currin2019give, tang2018awareness}. Research has shown that informal coordination relies on shared understanding and communication between caregivers, often mediated through tools such as calendars, pillboxes and digital messaging systems \cite{yamashita2018information, renyi2022uncovering}. These works point to the socio-technical nature of providing informal care, particularly in its reliance on effective communication between members of an informal network. 

\textcolor{black}{While providing valuable formative insights}, these studies often treat informal caregiving as a general category, without fully accounting for the specific care needs and relational nuances \textcolor{black}{of cognitive diseases such as MCI}. The assumption that informal networks can seamlessly take on caregiving roles \textcolor{black}{obscures} critical gaps, such as the lack of coordination tools that reflect the cared-for individual’s changing cognitive and functional abilites, or care roles shaped by cultural expectations that may not align with the diagnosed individual’s needs \cite{gutierrez2017takes}. These \textcolor{black}{gaps} underscore the need for an in-depth and grounded understanding of the provisions of informal care \textcolor{black}{for conditions such as }MCI\textcolor{black}{, which we discuss next.}

\subsection{\textcolor{black}{Care Coordination for MCI}}

Given that MCI is marked by mild cognitive decline with little to no associated physical decline, those diagnosed with it are often physically and cognitively able to go about much of their daily activities independently \cite{chandler2019comparative, albert2011diagnosis}. However, as the condition progresses, older adults with MCI face increasing difficulty managing complex instrumental activities of daily living (IADLs)\textcolor{black}{\footnote{\textcolor{black}{The National Institute on Aging (NIA) defines Instrumental Activities of Daily Living (IADLs) as complex tasks necessary for managing a household and living independently within the community. These activities require more complex cognitive skills, such as planning, decision-making, and organizational skills \cite{pashmdarfard2020assessment}.}}} such as medication adherence, financial planning, meal coordination and scheduling. Research has shown that informal CPs, particularly spouses, play a key role in supporting the older adult with MCI for such activities. A key aspect of this dynamic is that the primary informal CP for older adults with MCI is typically a spouse, who is themselves an older adult experiencing the typical cognitive and physical challenges that accompany aging \cite{madjaroff2017narratives, lydon2022integrative}. The needs of these CPs are unique not only in their relationship to the person diagnosed with MCI and the need to support them through their ever-changing capacity for autonomy, but also in their ability to manage the often sudden responsibility of becoming a caretaker, a role which they may feel ill prepared and equipped to take on \cite{adams2006transition, austrom2009long}. Spousal caregivers, in particular, may find themselves navigating unfamiliar responsibilities and emotional labor without formal training or systemic support \cite{dean2012living}. Their caregiving is not only task-oriented but also relational, shaped by decades of shared routines, tacit knowledge and deep interpersonal understanding.

\textcolor{black}{Since MCI is a diagnosis with a high incidence of conversion to the various forms of dementia, including Alzheimer’s Disease \cite{fischer2007conversion}, the coordination of informal care for MCI is complicated by the spectral nature and progression of the disease, which is unique \textcolor{black}{to} every individual \cite{busse2006mild, anderson2019state, hedman2013patterns}. Across \textcolor{black}{those} diagnosed, MCI can manifest  as a range of cognitive impairments (in memory, executive function, visuospatial skills or communication) and functional limitations that decline at varied rates and time scales, complicating the nature and means of \textcolor{black}{longitudinal} care coordination. This highly individual (and often unpredictable) nature of disease progression necessitates informal care mechanisms to constantly adapt to the changing abilities of older adults. This adaptation, in limited form, has been explored through the use of novel interventions intended to complement the coordination efforts of informal care partners \cite{mathur2022collaborative, zubatiy2021empowering}, where the technological system acted as a "stable support" in the face of uncertainty. Beyond the use of technologies, social support-based interventions have also proven beneficial \textcolor{black}{to} CPs \cite{lydon2022integrative, dean2012living}. However, research has also shown how many older couples dealing with an MCI diagnosis hesitate, and often refrain from seeking external help because of the adverse perception that the subtle cognitive decline of MCI is not as “serious” as dementia or other chronic illnesses \cite{madjaroff2017narratives, adams2006transition}, further highlighting the critical need to examine their in-home practices for care.} 

\subsection{Cognitive Decline and the Role of Technology}

A \textcolor{black}{substantial} body of work across HCI and CSCW has explored technologies to support care for people with dementia. These studies offer valuable insights into how digital tools can assist with memory recall \cite{alm2002designing}, task sequencing \cite{zaccarelli2013computer}, social connection \cite{kuwahara2006networked}, and caregiver coordination \cite{johnson2020roles, devries2019impact}. These works \textcolor{black}{also} highlight the role of assistive technological support for this stage of cognitive decline, leading to important design guidelines\textcolor{black}{. H}owever, such works have largely focused on older adults with moderate to severe cognitive decline, often at stages when \textcolor{black}{independence} has already diminished and caregiving becomes more directive and custodial. In contrast, relatively little research has been done on the provision and coordination of care for \textcolor{black}{the more subtle and progressive cognitive and functional decline associated with} MCI. Existing work has addressed various individual facets: differentiating MCI from normal aging and dementia \cite{lydon2022integrative}, capturing the lived experiences of caregivers \cite{lara2019functional}, or describing care for older adults with fluctuating \textcolor{black}{capacity for independence} \cite{corbin1985managing, lazar2018negotiating, schurgin2021isolation}. Other work has examined preventive technological interventions to delay MCI progression \cite{yang2022magic}, or highlight gaps in analog and digital coordination mechanisms \cite{smriti2024emotion}. Yet, these individual lines of inquiry often remain siloed, either focusing narrowly on the individual (person with MCI or caregiver), the disease trajectory (prevention or detection), or specific technology interventions, without examining how these elements converge within the day-to-day social practice of informal care coordination. This highlights an underexplored gap in the cohesive understanding of the socio-technical ecology of MCI care: how people, tools, relationships, and routines interact and evolve in response to early-stage cognitive decline.

A few recent studies in HCI and CSCW have begun to explore this intersection. Mathur et al\textcolor{black}{.} and Zubatiy et al\textcolor{black}{.}, for example, use participatory design methods and technology-focused interactions (\textcolor{black}{via an} AI-based system) with both the diagnosed individuals and their informal CPs to understand how technologies can support in managing everyday routines \cite{zubatiy2021empowering, zubatiy2023don, mathur2022collaborative}. While such works offer valuable formative insights, they tend to emphasize the design or evaluation of specific interventions, often within constrained domains (such as digital reminder systems or shared task lists), falling short of articulating the full complexity and variability of how MCI care is actually performed and coordinated across \textcolor{black}{everyday} contexts. 

In our work, we address \textcolor{black}{this underexplored gap} by offering a qualitative investigation of informal care coordination for individuals with MCI \textcolor{black}{that seeks to gain a comprehensive picture of this care coordination context as it unfolds in everyday lives and over time}. Our focus is not on a specific technology or isolated behavior, but on the broader landscape of people, activities, and tools involved in everyday care coordination. We investigate \textcolor{black}{these} realities of informal care coordination by asking: What coordination activities are performed, and how are they shaped by MCI-specific needs? Who participates in this coordination, and how do their roles evolve? What socio-technical assemblages, including tools, routines and communication strategies, mediate these care practices? By centering the lived experiences of both older adults with MCI and their informal CPs, we uncover how care coordination is not solely a matter of logistical organization, but also a deeply relational process shaped by negotiation with technological tools and evolving roles. We argue that MCI represents a critical transitional window: a time when individuals still retain significant agency and adaptability, offering a unique opportunity to introduce supportive technologies. Our contributions are \textcolor{black}{thus} twofold. \textcolor{black}{Empirically, we offer rich qualitative insights, grounded in interviews with diverse care networks, that characterize informal care coordination for MCI as orchestrated by a primary care partner. Conceptually,} we present a thematic \textcolor{black}{analysis} that captures the fluid and collaborative nature of MCI care coordination and surfaces actionable design implications for future socio-technical systems. In doing so, we aim to reframe MCI care coordination not as a simplified version of dementia care, but as a distinct and underexplored design space, one that demands systems capable of scaffolding autonomy, adapting to shifting capacities, and fostering collaborative caregiving practices.

\section{Study Method}

To examine the \textcolor{black}{informal care coordination practices} of older adults with MCI and their CPs, we conducted hour-long, semi-structured interviews with broadly worded questions about day-to-day supports, \textcolor{black}{with specific questions to understand the types of daily activities and routines that require support, the approaches and tools being used for coordination, and the individuals involved. Given the sparse previous work specifically focusing on older adults with MCI and their care partners, the interview methodology was deemed as most appropriate to gain a broad overview of their lived experience, and to identify future opportunities for more targeted and prototype design-based inquiry}. Interviews were recorded, transcribed, and analyzed qualitatively \textcolor{black}{to extract key themes}. \textcolor{black}{The recruitment, protocols, and data management followed IRB-approved policies to ensure safety, confidentiality, and voluntary participation.}

\subsection{Participant Recruitment}

\textcolor{black}{This research was conducted in partnership with a clinical program that supports people with MCI and their caregivers through a year-long curriculum addressing lifestyle, functional independence, social support, and cognitive strategies.} People diagnosed with MCI and their informal caregivers progress through this program as a pair, referred to in the program as a \textbf{dyad}. \textcolor{black}{The program refers to individuals with MCI as “members”, not patients, and calls their caregivers “care partners” to signify a mutual partnership and an active commitment to each other.} For our purposes, we will use the term person with MCI (pwMCI) in lieu of the term “member”, \textcolor{black}{and refer to the care partners as \textbf{CPs}, wherever applicable.} All participants recruited for this study were either current or past participants in this program at the time that they were interviewed. \textcolor{black}{In total, 7 pwMCI and 11 CPs participated in the interviews; however, one interview transcript was unavailable for analysis, and the results therefore reflect data from 6 pwMCI and 10 CPs.} \textcolor{black}{Not all interviews in our study included a pwMCI because, given the varying levels of cognitive decline presented by an MCI diagnosis, some CPs decided that it would be too cognitively taxing for the pwMCI to participate in the interview. However, we strived to include pwMCI, whenever appropriate, following prior research into informal care coordination that fostered richer discussion and discovery about shared daily routines when CPs and pwMCIs \textcolor{black}{were} interviewed together \cite{mathur2025sometimes}.} Demographic details about the participants can be found in Table 1. 

\textcolor{black}{Participants were recruited from the clinical research program through in-person outreach and email communication. In-person recruitment took place at the facility on \textcolor{black}{days when programming was offered}, where, with staff permission, the \textcolor{black}{primary} author introduced the study to CPs and members and collected contact information from those interested. These individuals were later emailed to arrange \textcolor{black}{virtual} interview times \textcolor{black}{(which we conducted via the video conferencing platform, Zoom)}. For email recruitment, the author also contacted program participants who had previously indicated interest in external studies, providing study details and an invitation to participate.}

\subsection{Qualitative Interviews}

We structured our interviews using the \textcolor{black}{three “who, what, how” dimensions} so that a holistic picture of participants’ daily routines and experiences could be gleaned. For our purposes, “who” pertains to the people involved in daily care coordination, whether they be other informal CPs or formal care contributors such as healthcare professionals. The “what” pertains to the activities that people with MCI and their CPs engage in which may require support and coordination, including those outside the home (e.g., appointments, errands, socializing), as well as at-home activities such as meal planning and preparation and home maintenance. The “how” pertains to any tools, both analog and digital in nature, that participants use or desire to support these daily activities. We specifically wanted to know whether participants tend to use the same tools in the same ways, and thus probed not only what tools were used but \textit{how} they were used.

A semi-structured format was used for the interviews to allow for guided conversation that both targeted certain high-level topics of interest but also gave participants the opportunity to express their individual perspectives on these topics. Two separate interview scripts were created: one script was used if both members of the dyad were present, and the other was used if only the CP was present. The script used \textcolor{black}{with} dyads included a \textcolor{black}{section of questions directed only at the CP, }to give them an opportunity to add \textcolor{black}{or expand on} anything without the pwMCI present (see Supplementary Files for scripts).

\begin{table}
    \centering
        \begin{tabular}{|p{2cm}|p{2.6cm}|p{3cm}|p{3cm}|p{2cm}|}
        \hline
         \textbf{\textcolor{black}{Dyad ID}}&  \textbf{Nature of relationship}& \textbf{Living Arrangement}& \textbf{pwMCI age}& \textbf{CP age}\\ \hline 
         CC01&  Spousal& Cohabiting& 83& 68 \\ \hline 
         CC02&  Parental& Remote& Not a participant& 59 \\ \hline 
         CC03&  Spousal& Cohabiting& Not a participant& 71 \\ \hline 
         CC04& Spousal& Cohabiting& 72& 68 \\ \hline
         CC05& Spousal& Cohabiting& 79& 72 \\ \hline
         CC06& Spousal& Cohabiting& 75& 74 \\ \hline
         CC08& Spousal& Cohabiting& 84& 80 \\ \hline
         CC09& Parental& Cohabiting& Not a participant& 43 \\ \hline
         CC10& Siblings& Cohabiting& Not a participant& 67 \\ \hline
         CC11& Spousal& Cohabiting& 84& 81 \\ \hline
    \end{tabular}
    \caption{Demographics for all interview participants}
\end{table}

\subsection{Data Analysis}

\textcolor{black}{Interview audio recordings, transcripts, and any identifying information were securely stored and de-identified during our analysis to protect participant privacy, in compliance with IRB-approved procedures.} We conducted the qualitative analysis \textcolor{black}{of interview transcripts in three distinct phases, which brought us from the initial “who, what, how” framing to our final cross-cutting thematic insights that we report in the Results.}  

\textcolor{black}{In the initial phase, the whole research team reviewed three full interview transcripts to identify a robust coding approach to capture the rich and varied content  of the interviews. This initial review led to the decision to analyze participant responses using three separate coding schemes, one each for activities, tools, and people. The activities scheme focused on  the “what” of care coordination, capturing  the tangible events, tasks, and activities that participants discussed in the context of day-to-day coordination. The tools scheme focused on  the “how” of care coordination, capturing the analog and digital products and services that participants use to organize and execute their daily activities. The people scheme focused on the “who” of care coordination, capturing the individuals who contribute  or are otherwise involved in the participants’ day-to-day activities, either in a professional or informal manner.}

\textcolor{black}{The second phase of analysis consisted of developing, refining, and applying the three inductive coding schemes to the interview transcripts. The first coding pass to each scheme was done by the primary author, and a second pass was done to each scheme by at least one additional coder. After this, any disagreements were flagged and resolved through group consensus. The coding approach involved flagging and adding each quote referencing a person, activity, or tool in one of three separate spreadsheets, along with any contextual details from the participants’ narratives. Each quote was also assigned a high-level category tag according to what was described (e.g., “driving” as an activity, “cell phone” as a tool, and “grandchild” as a person), as well as any sub-category tags, as described for each coding scheme below.}

\textcolor{black}{We started with \textbf{activities}, because we wanted to initially understand the targets of care coordination, and use those as anchors for insights about what tools were used, how, and by whom. The full list of activities was categorized into high-level codes, which captured both specific activities like managing appointments or socializing, as well as more complex activities such as planning for the future and navigating technology. Activities were also tagged with  standard categories often used to measure tasks that a typical person engages in on a day-to-day basis: Activities of Daily Living (ADLs, sometimes referred to as Basic ADLs or BADLs), Instrumental Activities of Daily Living (IADLs) \cite{pashmdarfard2020assessment}, and Extended Activities of Daily Living (EADLs) \cite{kim2025evaluation}.  Our final set of activity codes captured} whether each activity quote had a positive or negative valence, wherein positive quotes were deemed facilitators of coordination and negative quotes were deemed barriers to coordination. Facilitators were factors that supported the completion of the activity (e.g., helping a pwMCI engage successfully in a task, such as remembering to take medication because the pill box was labeled and there was a reminder on the home calendar), whereas barriers were factors that hindered the completion of the activity (e.g., the pwMCI failing to remember how to complete a task, such as scheduling an appointment via phone, which led to the CP having to complete the task for them). 

\textcolor{black}{While coding activities, we noted that participants naturally referenced the \textbf{tools} used to support the activity \textcolor{black}{as they described it}. Thus, we selected all activity quotes that mentioned tools and used them as a starting point to develop our codes for   tools, focusing on two main dimensions: objects and functions. The object codes captured broad categories of tools of similar form  (e.g., paper calendars, voice assistant devices, cell phones) while the function codes captured how a given tool could be used for a variety of purposes (e.g., a cell phone can have a planning/communication function and a scheduling/reminder function). This two-dimension\textcolor{black}{al} coding system allowed us to accurately track the multiple nuanced and idiosyncratic ways that different participants would go about using the same tools.}

\textcolor{black}{Finally, we developed a coding scheme to identify the \textbf{people} involved in informal care: individuals, dyads, and networks. The individual code captured activities and tools that specifically benefited the self, either the CP or the pwMCI (e.g., a CP organizing their own calendar or a pwMCI tracking their own medication intake). The dyad code, which was most common, captured coordination between the CP and pwMCI (e.g., organizing a shared calendar and planning social activities). The network code captured coordination that involved members of the informal care network outside of the dyad, including children, grandchildren, friends, neighbors, and hired support.}

\textcolor{black}{The third and final phase of qualitative analysis focused on identifying cross-cutting themes across the activity, tool, and people codes. This involved first the primary author taking the individual coding schemes and identifying relationships between them, such as when the mention of an activity included both the use of at least one tool and the involvement of at least one person outside the dyad. The primary author \textcolor{black}{then} created \textcolor{black}{a} relational mapping structure in an online design tool called FigJam\textcolor{black}{\footnote{\textcolor{black}{FigJam is a collaborative, online whiteboarding tool developed by Figma, designed for real-time team brainstorming, ideation and project planning, available at https://www.figma.com/figjam/}}}, which helped to visualize and capture the complex interrelationships among the  coding schemes and allowed for effective presentation and discussion of these relationships  with the larger research team. The results of this mapping used to facilitate this discussion are represented by Table 2 in the Supplementary Files.} 

\textcolor{black}{Using this approach of analyzing the relationships across the three dimensions as a team  allowed us to more clearly see how informal care coordination is instantiated  on a day-to-day basis, and to capture the breadth and depth of this unique coordination context. For example, our primary theme, characterizing CPs as care orchestrators, derived from an emergent pattern in our data wherein CPs planned activities that the dyad would engage in and introduced analog and digital tools as needed to help support the pwMCI in participating in the activity, as well as to allow other members of the care network to participate.}

\section{Results}
Our qualitative analysis yielded \textcolor{black}{three cross-cutting} themes that serve as the common thread linking all participants’ experiences \textcolor{black}{across the what, how, and who of informal care coordination \textcolor{black}{for} MCI. We intentionally did not structure our results around the “who, what, how” dimensions because once we analyzed our data using this methodological approach described above, it became clear that the important insights derived from our data sit at the intersection of the three coding schemes. Presenting our results by cross-cutting theme as we have below more fully captures the depth and complexity of insights that our methods revealed. Our three themes are as follows:}
{\begin{enumerate}
    \item \textcolor{black}{Informal CPs as orchestrators of care activities;}
    \item \textcolor{black}{Managing the transitional nature of MCI and uncertainty about future cognitive decline;} 
    \item \textcolor{black}{Expanding and calibrating the caregiving network in anticipation of future care needs.} 
\end{enumerate}
}

\textcolor{black}{We include representative quotes from our interviews to contextualize and support each theme, wherever relevant. Quotes are labeled using the notation CCP/M\#\#, where \#\# refers to the dyad ID and P/M indicates their role within the dyad: as Care Partner or \textcolor{black}{Care} Member.}

\subsection{Theme 1: \textcolor{black}{Informal CPs as Orchestrators of Care Activities}}
A common theme that emerged was the need for care partners to \textcolor{black}{take on a central} orchestrat\textcolor{black}{ion role, whereby they would not only support their loved one with MCI, but also mediate the involvement of other members of the care network in daily activities. As discussed below,} this orchestration could take on both direct and indirect forms. \textcolor{black}{In some situations,} the CPs were directly involved in the planning and execution of tasks \textcolor{black}{to support the pwMCI. In others, their orchestration was indirect, often taking the form of} intangible \textcolor{black}{emotional} considerations \textcolor{black}{when deciding how to support the pwMCI in order to maximize their sense of autonomy and independence.} Another key element of this orchestration was that CPs had to take on different mediation roles within their \textcolor{black}{care} networks, \textcolor{black}{creating} a centralized coordination pipeline \textcolor{black}{by which} members of the care network were brought into the process through collaboration with the primary CP.

\subsubsection{Direct and indirect orchestration of activities within the dyad}

\textcolor{black}{Direct orchestration captures the fact that often, the CP is directly making arrangements for supports, such as scheduling pwMCI's appointments to coincide with theirs, driving somewhere together to make sure the pwMCI doesn’t get lost, and creating visual reminders for upcoming tasks and activities. In these instances, the need for an activity was clear and the effect was tangible to the pwMCI:}
\\

\textit{CCP05:“Our regular doctor that we see, we always set it up with two appointments right next to each other and we go together. So we go into the room together. She does both of our workups and stuff. And then we schedule for the next time."}
\\

\textcolor{black}{This demonstrates a system of planning a recurring event in a way that is efficient for the dyad and comfortable for the pwMCI. CCP05 \textcolor{black}{also} referenced the use of a digital calendar as the dyad’s means of tracking the scheduled appointments, which highlights the cross-cutting nature of this activity across tools and people. Direct orchestration of activities also cut across the use of analog tools:}
\\

\textit{CCP10: “I have a whiteboard by the door where we put some tasks that we wanna see before we're going out the door. Something important that has to be done that week…it's facing the kitchen table and [we] tend to put things that are more imminent and important to get done. So, if we're expecting that…we have to have something ready for a delivery man to pick up, it's put there. Or if there's a specific important task that we wanna have a lot of reminders that we have to do, it goes there.”}
\\

\textcolor{black}{In contrast, indirect orchestration represents the "behind the scenes" planning and arranging of scaffolds by the CP in order to enable the pwMCI to remain engaged in daily activities and to complete them at their current level of independence. This included taking into account the pwMCI's psychological and emotional needs, such as their need to feel that they were still contributing to the household or still in charge of activities that they had always taken care of pre-diagnosis. For tasks that the pwMCI was no longer capable of doing independently and that the CP could technically complete more efficiently on their own, the CPs would instead arrange needed supports to enable the pwMCI to complete them independently, or with minimal assistance from the CP:}
\\

\textit{CCP11: "I know that there has to be an arrangement…And I want it to be for him and with him…I'm gonna start trying with notes…for communication...like with small tasks, pay the water bill, pay the light bill, pay American Express…And then following up to make sure that things are done and get him in the habit of doing it."}
\\

\textit{CCM11: “The [Alexa-enabled device] in the den tells me to take out the trash around eight o'clock every night.”} 
\\

\textit{CCP05: “We used to write lists of things…Now what we do is I'll say it's time for groceries. And I'll tell [Alexa] that she needs to clear the old grocery list. And then CCM05 will stand in the kitchen and he'll look around at things…He does not like to tell Alexa what to put on the list...so I'll sit there and he'll say, we need milk. And I'll say, Alexa…add milk to the shopping list. And then we'll go through that whole list…And then at the end I'll tell her to go ahead and print the shopping list. And then we use that shopping list to go get groceries.”}
\\

\textcolor{black}{These examples, while varied in nature, represent different ways that CPs found to support  the pwMCI through physical scaffolding and emotional attunement to their need for independence. This suggests that using indirect orchestration was not merely a means to achieve a task more efficiently through collaboration or prep work, but also a way to enable the pwMCI to maintain their sense of autonomy and to cognitively engage them for purposes of enrichment and stimulation.}

\textcolor{black}{Similarly, if a pwMCI was capable of doing something independently, but the CP wanted a means of checking in on that activity, they found different ways of setting up technological supports for remotely checking in with the pwMCI. These supports were also context-dependent and considerate of the forms of independence that the various pwMCI sought, such as when a pwMCI enjoyed going on walks alone but couldn’t always find the way home:}
\\

\textcolor{black}{\textit{CCP09: “When she got lost in the neighborhood, I was like, okay so she can’t be left alone...during the day, she needs somebody there, and the iWatch became very, very important...it was really her lifeline to me and my lifeline to her.”}}
\\ 

\textcolor{black}{Or when the pwMCI wanted to keep an independent living situation despite needing daily reminders from the CP:}
\\

\textcolor{black}{\textit{CCP02: “I was able to place their iPad in a location where I can leave it, stationed and open and plugged in because they're not gonna remember to charge it…and I FaceTime through the iPad and I almost do it daily. And they both know how to open it up and see me.”}}
\\

\textcolor{black}{Here, technology served as a way to give CPs peace of mind while ensuring the continued means of independence that the pwMCI were accustomed to and therefore benefited from continuing. Whether that be going on walks alone (iWatch) or living alone (iPad/FaceTime), remote check-ins via aptly placed devices proved vital for providing the peace of mind CPs sought while simultaneously ensuring the independence and quality of life that pwMCI were accustomed to.}

\textcolor{black}{Indirect orchestration also took the form of coordinating assistance \textcolor{black}{from} care \textcolor{black}{providers} outside the dyad to maintain  a baseline day-to-day experience that the pwMCI had become accustomed to prior to their diagnosis. \textcolor{black}{This} generally came about as a result of CPs wanting supports for the pwMCI that they could not execute themselves and therefore had to rely on others to provide, though they maintained responsibility for planning and coordinating these supports ahead of time:}
\\ 

\textit{CCP02:“We moved them to an apartment, independent living in an assisted living facility…someone…
comes every week to clean and two meals a day and…a number of activities. But all that was coordinated by us. They are no longer capable of doing that. And I think it's because my mom [was] the one who always did it, and she's the one whose memory is failing. My father always [depended] on my mom, so my mom's memory failing means my father now needs help. So we are pretty much doing mom's job for that, everything that you can imagine is done behind the scenes.”}
\\

\textcolor{black}{Although this CP describes the tangible\textcolor{black}{,} direct action of helping her parents move into an assisted living facility, what this explanation conveys is the indirect orchestration considerations that underlie such a move. This scenario suggests that, in order for her parents to live independently in a way that they can manage day-to-day, the CP had to be deeply involved in ways that were not directly apparent to the pwMCI. This and all examples of indirect orchestration allow for significant maintenance of quality of life due to the emotional awareness that CPs have for the delicate balance between a pwMCI’s challenges and their need for autonomy and self empowerment. Furthermore, the utilization of both analog and digital tools across direct and indirect orchestration scenarios indicates a pattern of use wherein tools are generally leveraged within the dyad as a means of planning and managing activities that involved both the pwMCI and the CP, suggesting that orchestration efforts provide a means for CPs to not only support the pwMCI, but to simultaneously support themselves.}

\subsubsection{Care partners as mediators between and across the broader care network}

\textcolor{black}{Another key aspect of orchestration was that  in addition to supporting the pwMCI themselves, the CPs acted as mediators between the pwMCI and their pre-existing formal and informal care network, facilitating their involvement in supporting the pwMCI. Often this included enabling the pwMCI to stay connected with grandchildren and other family members, as illustrated by the following:}
\\

\textit{CCP01: “We're constantly getting pictures of our grandchildren and their activities, things that actually are of interest to him, but he doesn't know it unless I say, ‘Oh, we got a picture. Go take a look at it,’ or even more annoying to me is that I have to be the secretary now because I'll get pictures…‘Oh can you send them to me?’ So that's like an additional step for me.”}
\\

\textit{CCP04: "He understands when I tell him [that] she [their daughter] is in pain or when she's...applied for a job that she didn't get. But I have to be…the mediator, if you will, between him and our children, because...now I have to say [\textcolor{black}{their} son] will call him, and [\textcolor{black}{their} daughter] and I just set up a weekly call that will be to me first, and then I'll bring it to CCM04."}
\\

Mediation also took place when a CP \textcolor{black}{made arrangements for others to watch over their loved one with MCI so that they could attend to personal matters outside the home, including ensuring that the outside care partner was aware and able to support the needs of the pwMCI:}
\\

\textit{CCP03: “I have surgery coming up for my eyes, two different surgeries and my baby sister…she's gonna come and ‘hang out’ with CCM03 while I'm gone...I can ask her to do that kind of stuff, but she's not in a position say to, pick him up if he falls or, that sort of thing.”}
\\

\textcolor{black}{Digital tools were common means by which the mediation function was accomplished, as they allowed for more robust coordination across time and space. However, as the following quote illustrates, even a reliance on digital tools did not obviate the need for additional labor \textcolor{black}{by} the CP to ensure that information is centralized and up to date for the common goal of providing support:}
\\

\textit{CCP02: “In a Google Doc…I put all the medications and my job was to update it. So every time someone took him to a doctor's appointment, they could open a Google document. When they ask what are his medications…they will see it. But if anything changed then I had the responsibility, even though I did not go to that appointment…to figure out how to keep it updated. So it required a little bit more, follow up on my siblings. Either training them, ‘you need to go in and update it,’ when that was not going to happen. Then it is about ‘you need to send me a text’…many times it never got updated. I'll be honest with you…when there's more than one [sibling], it's very hard.”}
\\

\textcolor{black}{Adult children CPs who were not co-habitating with their parent with MCI or whose work responsibilities took them outside the home had to additionally coordinate and mediate communication among non-family caretakers outside the informal care network, such as home health aides who were brought in to assist the pwMCI. Interestingly, as one CP related, analog tools were much easier than digital tools to facilitate communication across multiple outside caretakers:}\\

\textit{CCP09:  “In terms of coordinating care, this diary became really crucial, especially as I started bringing in...outside people to help watch her when I wasn't there. But then it was still me going through the forms and going, okay…What did she do today? Is her mood worse today? Is that because of this medicine or is it because of this other thing?...It is easier when I have the big notebook and I can just go flip, flip. Look at that, flip, flip over here. In the physical form, it was helpful to notice trends…the handwritten was easier across the board for both the caretakers to fill out…they could just go in and check.”}
\\

\textcolor{black}{Similar to the example above, the primary CP, in his orchestrator role, still maintained the responsibility for setting up a system that worked for the care network and \textcolor{black}{keeping} the “big picture” of his mother’s health in mind across multiple informants.}

\textcolor{black}{In these examples of mediation, it’s clear that CPs coordinated care in ways that suited their unique mix of informal care network members and  outside hired support to ensure that the needs of the pwMCI were being met. This was demonstrably a non-trivial amount of labor, indicating that regardless of who is involved in the coordination of care, what tools are used to do so, and the activity being coordinated, mediation is a complex orchestration task best achieved with one lead care orchestrator. This task falls on the primary CP who typically knows the pwMCI best of all and can therefore most effectively balance their own needs and the needs of the pwMCI. This mediation also differs from orchestration within the dyad as mentioned in Section 4.1.1 above in that the tools used, whether analog or digital, generally serve as means of tracking outside support rather than planning for it. This suggests that when using tools for mediation with the greater care network, CPs are doing so in a way that helps them manage the outside support, rather than simply collaborate on the work that needs to be done.}

\subsection{Theme 2: \textcolor{black}{Managing the Transitional Nature of MCI and Uncertainty About Future Cognitive Decline}}

\textcolor{black}{The second theme regarding coordination of informal care for individuals with MCI centered around the progressive nature of the disease and the resulting uncertainty about the anticipated future care needs of the pwMCI. Informal care coordination for MCI represents a “moving target” whereby care partners must be aware of and monitor the pwMCI’s declining cognitive and functional skills, and continually adjust and calibrate the supports, tools being used, and involvement of others to account for these changes.}

\subsubsection{\textcolor{black}{Learning how to coordinate for MCI at each progressive stage of care.}}

\textcolor{black}{Although the dyads interviewed for this study shared a common experience of coordinating around a diagnosis of MCI, each provided a unique perspective and set of experiences regarding the ways \textcolor{black}{in which} an MCI diagnosis impacted their lives. These perspectives were heavily shaped by how long the participants had been on the care coordination journey at the time of being interviewed. We identified this spectrum of experiences as a staged progression of the diagnosis, from early-stage to mid-stage and finally late-stage when the diagnosed individual is transitioning into dementia. These stages are also supported by prior work [21]. Despite the unique lives and circumstances of each individual diagnosed with MCI, there were stage-by-stage similarities in how CPs described the amount and type of supports the pwMCI needed to engage in daily activities.}

\textcolor{black}{With respect to the experiences of someone whose diagnosis was recent at the time of being interviewed, one pwMCI noted:}
\\

\textit{CCM08: “The situation is that I'm not completely helpless. I don't get lost in stores and stuff like that.”}
\\

\textcolor{black}{This self-reflection suggests that the early stage of diagnosis is marked by an awareness from the pwMCI that cognitive decline beyond the normal signs of aging has started to set in. This specific dyad also went on to discuss how the pwMCI had started taking measures to self-support against his increased forgetfulness by creating a handwritten system for tracking his medication intake. Across our participants, this kind of behavior was unique to those whose MCI diagnosis was recent at the time of the interview.} 
\textcolor{black}{Individuals in the middle stage of the disease were generally less aware of their limitations, \textcolor{black}{causing CPs to speak more on behalf of the pwMCI,} as illustrated by the following:}
\\

\textit{CCP01: “Despite what he says, that he wants me to do things so that I learn how to do them, that really isn't true because I'm obviously a very capable woman…Someone who always took care of the house and then they stop doing it, it's not laziness…that's sort of a rationalization…the slowing down of your own capacity.”}
\\

\textcolor{black}{Here, the CP’s reflection of the fact that the pwMCI finds reasons for not doing tasks that he used to do routinely suggests that there may be an element of denial in the middle stages of the diagnosis. An acknowledgement of uncompleted work does not translate to a change in behavior, as a declining cognitive capacity lessens the ability for pwMCI to go about their day-to-day activities autonomously. In response to this increasing gap between physical and cognitive capacity,} dyads \textcolor{black}{at this stage of the journey} found creative technological solutions to support the pwMCI through tasks affected by their declining cognition:
\\

\textit{CCP05: “He's not big on wanting to enter things in, and as this disease has progressed, he's lost the ability to spell. So now…I put a notes document or app on his phone that he can just speak into the app and then it spells it.”}
\\

\textcolor{black}{Tactical measures such as this provided a way to meet pwMCI where they were in terms of ability and motivation at a given stage of their diagnosis. The idiosyncrasies present in each individual’s progression indicated that in order for a supportive measure to be effective, it had to be responsive to their present state of self, regardless of their past interests, skills, and pursuits. For example, when contrasting this experience from CCP05 to CCP01’s expression of frustration as having to act as a sort of phone secretary for her husband [Section 4.1.2], there is a prominent difference in the way these two pwMCI interact with their phones for the purposes of asynchronous communication. The approaches used by each CP were significantly different as a result, pointing to the importance of understanding someone’s unique care needs despite similarities due to a shared stage of progression.}

\textcolor{black}{If the pwMCI was in the late stages of \textcolor{black}{the} disease, they were unable to do much independently and often not aware of their limitations, which created a \textcolor{black}{significant} practical and psychological burden on CPs who found themselves needing to step in \textcolor{black}{during} almost all aspects of daily activities:}
\\

\textit{CCP02: “My mom has been MCI for over 12 years. As she progresses in MCI…and her memory starts to decline, then I am more and more involved. I don't think a day goes by that I don't talk to them or I don't see them at this point.”}
\\

\textcolor{black}{In this case, despite the continued and concentrated efforts of the CP, the diagnosis continued to progress over the course of many years, creating new norms for the dyad along the way. In Sections 4.1.1 and 4.1.2, CCP02 recounted various ways that such an evolution has brought this dyad’s informal care network to their current experience, involving multi-person asynchronous tracking and remote management of semi-independent living. These practices were developed over the course of this decade-plus-long endeavor, pointing to the complexities apparent in managing care coordination longitudinally.}

\textcolor{black}{\textcolor{black}{Collectively, t}he pwMCI we interviewed were experiencing cognitive decline at different rates and were at varying timepoints since their initial diagnosis. Those who were interviewed closer to the time they were diagnosed characterized their coordination experiences much differently than those who had received the diagnosis long ago, with a recognizable shift in decreased capacity and increased need for supports over the course of the disease. An awareness of care coordination needs as a longitudinal phenomenon that changes over time is essential to developing appropriate tools and methods for supporting pwMCI and their CPs through these transitions.}
\\

\subsubsection{\textcolor{black}{Navigating uncertainty about future care needs.}}

Due to the progressive nature of MCI, CPs expressed uncertainty about the future, both in terms of the care decisions they were currently making and anticipated evolving care needs. The most common reason for this uncertainty was their lack of knowledge or confidence in making decisions now that would maximize their \textcolor{black}{own} and their loved one’s quality of life later on. One example of uncertainty around taking actions now that would support future needs concerned the potential for needing to move to a more supported living situation:
\\

\textit{CCP08: “Of course we’re dealing with the question that everyone deals with, which is, do we stay or do we go in our house? And we went through this exercise last spring, seriously thinking about selling the house and moving to some sort of independent living residential facility. And we just couldn’t pull the trigger on that. So here we are, living in a house we’ve lived in for 50 years with all the stuff and realizing that at some point, we will have to leave.”}
\\

\textcolor{black}{This couple recognizes that although change is imminent and important for their future needs, they find themselves unable to commit to that change. The mention of living in a home for 50 years implies a resistance to departing from a comfortable and familiar situation, but it also suggests confusion about the best way forward. They lack confidence in their ability to decide whether to pursue a new residence or added support at their current residence, which is challenging as these kinds of decisions are deeply personal and complex, yet dyads feel a sense of urgency to address them.}

\textcolor{black}{The progressive nature of MCI also factored into how CPs contemplated the roles that other informal care partners may play in supporting the pwMCI at a future stage of care. For example, CCP06 spoke of a previous heart attack he had that required a week-long hospital stay where his daughter stayed with CCM06. This arrangement caused no issues but raised questions for CCP06 about the feasibility of a similar situation should it happen at a later stage in the diagnosis:}
\\

\textcolor{black}{\textit{CCP06: "If you’re early stages...it takes some adapting to, but there’s not a lot of major concerns…But if they’re in a later stage, then I could see where [there] would be major concerns about...keeping up with in-house care or scheduling appointments.”}}
\\

With respect to tools, uncertainty about the future manifested through participants postulating about ways that technology could support them now and later on as the pwMCI’s cognitive symptoms worsened:
\\

\textit{CCP01: “Auditory prompts are better than visual prompts...I could see in CCM01's case that if he became more forgetful…and I wanted to remind him to do something, if I had a way to communicate with a speaker in the house and say, ‘take your medicine,’ or, ‘you have a doctor's appointment at three o'clock today.\textcolor{black}{'}”}
\\

\textcolor{black}{Participants showed strong foresight and awareness into the kinds of technical supports they would find useful as the MCI diagnosis progressed, but filling the gap between recognizing the value in new tools and implementing those tools in ways that addressed the dyads' lifestyles, preferences, and day-to-day needs proved challenging. This was true regardless of tech literacy within the dyad or from members of the larger informal care network, due to the boilerplate nature of many solutions available on the market today.}

In other instances \textcolor{black}{pertaining to the usage (or lack thereof) of tools,} pwMCI would express their dislike or lack of desire for technology altogether, explaining that in their perception, technology will not improve their quality of life going forward:
\\

\textit{CCM11: “Only reason I need a phone is to call her or to receive a call from her. Other than that, I don't need a phone."}
\\

\textcolor{black}{Overall,} the lack of tools that aligned with their current needs meant that, even  technology that was readily accessible to these dyads was perceived as not robust or intuitive enough to be leveraged in ways that supported their day-to-day \textcolor{black}{activities}. As a result, the idea of using  technology to support future changes in abilities and needs was perceived as potentially futile. These experiences collectively suggest not only practical challenges, but an emotional weight experienced by CPs. \textcolor{black}{The uncertainty} was brought about by not knowing what the future would bring, whether they were equipped to handle it, and what supports they could implement now, technological or otherwise, \textcolor{black}{to} help anticipate \textcolor{black}{and reliably support changing} future needs \textcolor{black}{over time. Most importantly, these sentiments were shared across all dyads, speaking to their prominence across the spectrum of MCI care coordination contexts}. 

\subsection{Theme 3: Expanding and Calibrating the Care\textcolor{black}{giving Network in Anticipation of Future Care Needs}}

\textcolor{black}{Related to \textcolor{black}{their} uncertainty about the future due to the progressive nature of MCI, CPs also expressed a desire to expand the dyad’s support and social network. This included involving others in supporting day-to-day activities, creating new opportunities for socialization, and expanding the care network to include hired help such as gig workers in anticipation of greater needs for support in the future. While above we discussed ways in which CPs mediate among current members of the care network, here we focus on their considerations and efforts around bringing new people into the network, as well as expansion and calibration efforts that stem from CPs fine\textcolor{black}{-}tuning the ways in which existing members of the care network provide support. This included, for example, the necessity of bringing in outside help if the pwMCI needed any sort of physical support in the future:}
\\

\textit{CCP04: “CCM04 is six foot four and cancer really weakened me…if he gets to a place where he needs help...I don't know if that means in-home. We could actually have somebody live with us.”}
\\

\textcolor{black}{The need for outside support pushed dyads to consider situations that previously felt foreign to them, demonstrating the lengths to which CPs would go to find the right kind of care for their loved one. This combined with CP awareness of their own preferences and abilities made for effective decision-making about care network expansion across our participants.}

\textcolor{black}{Socialization was another commonly sought form of new outside help. Traditional forms of social interaction were not always a manageable form of engagement to orchestrate, so CPs would find creative ways to create social interaction via hired gig workers:}
\\

\textit{CCP03: “We have a guy who cleans our house once a week, but the reason we want him, I want him to keep doing that, was that CCM03 needed to have somebody around if I went to the store, or if I went over to a friend's house for a couple of hours. So that's a very safe thing, but…we're getting to a point where we are gonna need to get more help.”}
\\

\textit{CCP08: “CCM08 is getting picked up to go to lunch with a friend, and the friend has a man that comes several times a week and helps him with things and drives him where he needs to go…if we ever needed that and could afford it, we certainly would look for someone.”}
\\

\textcolor{black}{These strategies were effective not only because they offered the support that CPs sought, but because that support came in the form of social enrichment for the pwMCI. Engaging the cognition of pwMCI in this way has been suggested to provide benefits related to quality of life \cite{lydon2022integrative, dean2012living} and it also serves the pragmatic purpose of providing respite to CPs from their care coordination responsibilities. Although both CCP03 and CCP08 note that this support will need to evolve over time, the considerations they express in exploring how to best support social enrichment opportunities for their spouses makes clear the acute awareness they have for which non-traditional means of support will best suit the dyad’s needs.}

\textcolor{black}{However,} despite CPs acknowledging the anticipated need for outside help, because they currently take on the lion share of care responsibilities, that concept felt foreign \textcolor{black}{for some participants}. Notably, this sentiment was also expressed by pwMCI, who often did not want someone other than their primary CP to be involved in their daily lives:  
\\

\textit{CCP04: “I was talking to our daughter yesterday, and she thinks I need to start bringing in help now before he gets to the point where he absolutely needs it. I'm not sure I agree, but I hear her and he was really not comfortable. He can't remember that obviously, but he was not comfortable when I needed help brought in.”}
\\

\textcolor{black}{The identification of actual or potential resistance from pwMCI put CPs at a crossroads about how to maximize both the support needs as well as the preferences of both members of the dyad. This led to strategizing about how to ease a pwMCI into new forms of outside support, which gives even further import to the value of unique and ad hoc forms of social engagement. Such context-aware approaches to outside support have the potential to serve as a first step toward broadening the care network via means that adequately consider the comfort and preferences of a given pwMCI while ensuring their continued success.}

CPs also noted the \textcolor{black}{logistical} challenges associated with bringing outside support into the care coordination network\textcolor{black}{, especially in instances where familial and other pre-existing social support was limited or otherwise not preferred}:
\\

\textit{CCP10: “I guess the worst-case scenario was if something happened to me in total that a professional group would have to come in and liquidate the contents of the house for her and sell the house for her. I'm not sure how that would be handled and I don't know if I have to get an attorney to determine something like that.”}
\\

\textcolor{black}{Technical challenges like this one speak to yet another layer of complexity embedded into the care network expansion process by someone who has grown accustomed to managing most if not all care coordination tasks themselves. Ultimately,} whether the need for future help would come about as a result of the CP no longer being able to provide care on their own, or the pwMCI’s level of need exceeding the CP’s capacity to provide care, every dyad expressed concern and uncertainty about future supports. \textcolor{black}{Inability or preferences not to rely on} family members, lack of outside resources, \textcolor{black}{financial limitations,} and lack of trust in anyone other than the primary CP meant that selecting the right kind of help became an ongoing source of confusion and concern. \textcolor{black}{Such dyad-specific nuances suggest that while there is no “best” or “correct” way to involve outside resources into the care coordination experience, there are approaches which can be utilized to identify and explore opportunities that suit individual needs and contexts.}

\section{Discussion}
Our analysis revealed insights about the nature of informal care coordination that both support and extend prior work. Prior research \textcolor{black}{has noted that spouses play a key role in supporting their partner with MCI in navigating daily routines. Because these care partners are also older adults, they find themselves in a unique position of dealing with the cognitive and physical challenges that accompany normal aging, while simultaneously supporting their spouse with MCI through their increasingly declining capacity for independence. In turn, this long-term, gradually expanding set of caretaking responsibilities, combined with a lack of support tools tailored to these caregiving tasks, brings with it mental, emotional, and physical challenges.} Our results corroborate these \textcolor{black}{prior findings and highlight both} the complexities inherent in providing robust care coordination, \textcolor{black}{as well as the myriad} considerations that CPs must derive tacitly as they learn about and navigate that role. This includes coordination in areas such as activities of daily living, social engagement and entertainment, healthcare needs, and \textcolor{black}{socio-emotional} needs, while CPs are supporting their own needs and wellbeing.

\textcolor{black}{Our work adds nuance to this picture by highlighting orchestration as a key mechanism by which care coordination is enacted by care partners.  We use this term to highlight the fact that care partners not only make arrangements, both directly and indirectly, to ensure that their loved one with MCI is supported through their day to day routines, but also mediate the involvement of other members of the extended care network in providing these supports. Thus, much like a musical conductor who interprets the score, sets the tempo, and cues musicians to unify their performance, so do primary care partners play the central role in identifying the types and levels of support needed, the means by which these supports are delivered, and the roles that family members and other care partners will play. Moreover, orchestration as a form of coordination also acknowledges that informal coordination practices involve a unique form of emotional awareness and consideration that come about in response to the needs presented by an MCI diagnosis. These include not only cognitive challenges but also socio-emotional needs, such as a desire for autonomy, independence, and maintaining quality of life. The result is a set of direct and indirect orchestration activities enacted by CPs that provide both practical  and socio\textcolor{black}{-}emotional benefit, in turn enabling pwMCI to continue to support themselves in ways that CPs deem feasible given their current cognitive state. We propose that this uniquely situated context necessitates the consideration of new technologies that address the orchestration needs of informal care for MCI.}

\subsection{Coordination as Orchestration Across \textcolor{black}{the} Care Networks}

\textcolor{black}{Care coordination is an umbrella term that encompasses both the formal means by which healthcare professionals collaborate to manage a patient's healthcare journey, as well as informal practices by which unpaid family or friends organize and manage care for a loved one, often alongside formal healthcare. With respect to Mild Cognitive Impairment, our work unpacks informal care coordination as an orchestration role taken on by a primary care partner, typically a spouse,  who takes on the main responsibility for arranging supports for the pwMCI, both within the dyad and across the care network. This orchestration requires significant effort, planning, and awareness that amounts to a complex array of compensatory supports that allow the pwMCI to maintain a high quality of life. This unique form of coordination, which we define through the act of direct and indirect orchestration, differs from other kinds of informal coordination. It is characterized by the “soft” coordination that happens when a care partner orchestrates their efforts around the cognitive and emotional needs of the pwMCI.}

\textcolor{black}{Other works explore the scaffolding that often happens in informal care scenarios \cite{schurgin2021isolation, zubatiy2021empowering, bhat2023we}. Zubatiy et al\textcolor{black}{.} discussed how care partners scaffolded the use of a Google Home device by a loved \textcolor{black}{one} with MCI by setting up alarms, reminders, and calendar events that the pwMCI could subsequently \textcolor{black}{engage with}. While our work similarly documented evidence of scaffolding by CP\textcolor{black}{s that was} intended to compensate for the cognitive challenges presented by MCI, we  additionally found that CPs orchestrate supports intended to empower and enrich the mental and emotional state of the pwMCI \textcolor{black}{by involving them} in tasks where their help was not necessarily needed but perceived beneficial as a form of cognitive stimulation\textcolor{black}{. An example of this was} when one CP \textcolor{black}{encouraged the pwMCI to start paying bills despite the CP having always done it previously, following up afterward to make sure they were paid properly [CC11, Section 4.1.1]}. This action demonstrated a keen awareness of the value of keeping the pwMCI involved in a task that he could no longer complete  alone but could manage if provided emotionally attuned structure.}

\textcolor{black}{\subsubsection{Informal orchestration as a gateway to formal support}
A key theme that emerged \textcolor{black}{in our results} with respect to informal care coordination in MCI was the central role that the primary CP, typically a spouse, played in orchestrating formal support from outside the dyad, \textcolor{black}{for themselves and their loved one with MCI, beyond their immediate care network.} This outward facing orchestration was previously identified in a large scale survey of informal caregiving in the US \cite{schurgin2021isolation} wherein CPs were found to be the conduit for the effective involvement of formal care workers, such as healthcare professionals. Our insights illustrate the breadth and complexity of this role in practice, highlighting the many ways in which informal orchestration of care often brings about \textcolor{black}{external} support. For example, our interviews revealed many instances of care partners arranging for someone to spend time with the pwMCI while the they were away from the house, such as \textcolor{black}{one} CP who continued to hire a cleaning person to watch over her husband \textcolor{black}{[CC03, Section 4.3]}, and another CP who was contemplating hiring someone to drive her husband around to run errands \textcolor{black}{[CC08, Section 4.3]}. Other examples include hiring financial planners to prepare documentation in anticipation of future care needs \textcolor{black}{[CC10, Section 4.3]}.
}

Importantly, these orchestration efforts were \textcolor{black}{both} logistical and emotionally attuned to pwMCI’s needs and expectations. When formal support deviated from established expectations, CPs worked to recalibrate and negotiate how support was introduced so that it felt familiar and non-threatening. For example, one \textcolor{black}{CP whose loved one expressed concerns with bringing \textcolor{black}{in} outside help} explained that they would only consider long-term professional support if it could be brought into their current home, and only if it was absolutely necessary \textcolor{black}{[CC04, Section 4.3]}. This dynamic underscores and highlights that the integration of formal support into informal networks is rarely seamless, and requires constant negotiation from CPs to preserve the pwMCI’s sense of autonomy. For designers of socio-technical systems, this suggests that in addition to facilitating communication with formal care providers, the design of care coordination technologies should also acknowledge and scaffold the hidden labor CPs perform in making formal care more approachable and acceptable to pwMCI.

\textcolor{black}{\subsubsection{Balancing day-to-day needs with involvement from outside the dyad}
In addition to serving as gateways to formal support, we also observed how CPs had to manage the delicate task of \textcolor{black}{facilitating and enabling the involvement  of other members of the informal care network  in the daily life of the pwMCI.} In doing so, they navigate complex social dynamics with family and friends who want to be involved but have their own prerogatives for doing so, or those who do not know how to be involved without significant scaffolding by the CP. For example, \textcolor{black}{one CP described the} inconvenience involved in bringing their siblings into the care network\textcolor{black}{,} highlight\textcolor{black}{ing} the effort required by CPs to manage expectations and priorities across the caregiving ecosystem \textcolor{black}{[CC02, Sections 4.1.2 and 4.3]}. \textcolor{black}{Another CP's} sister faced \textcolor{black}{potential} difficulties in caring for \textcolor{black}{the CP's} husband as a result of \textcolor{black}{physical} limitations \textcolor{black}{[CC03, Section 4.1.2]}. These experiences highlight how CPs must continually balance competing expectations: accommodating the preferences of those eager to participate, while also lowering the barriers for those whose involvement may require additional support.}  

\textcolor{black}{Every CP identified elements of their lived experiences where support from other members of their informal network was important for their quality of life. Prior work has shown that informal caregivers often negotiate the involvement of others based on a need\textcolor{black}{s} and cost/benefit analysis \cite{schurgin2021isolation}. Our work exposes a greater nuance to this reality, and sheds light on the nature of work involved in this coordination process. We highlight how CPs are required to engage in consistent, thoughtful, and careful planning to orchestrate support for pwMCI, regardless of how much the severity of MCI impacts their ability to engage in day-to-day activities.}

\textcolor{black}{The recognition of this distributed caregiving also speaks to prior work around shared goals among informal caregivers, and the resulting communication breakdowns. In a collaborative caregiving network, such breakdowns and mismatched expectations can create friction, miscommunication, and even conflict \cite{consolvo2004technology, currin2019give, tang2018awareness}. Prior work in HCI and CSCW focusing on care for advanced dementia also noted similar frictions, particularly around communication across remotely located caregivers \cite{johnson2020roles, devries2019impact}. Our findings add another dimension to this body of work, one that is centered on the MCI experience that often precedes dementia. We highlight how similar coordination tensions emerge well before the onset of dementia, in the context of MCI, where individuals still exercise significant autonomy but CPs must already anticipate and plan for an uncertain trajectory of decline, shaping their use of technological tools that may support them through it. This suggests that future systems should facilitate task distribution, and also explicitly scaffold negotiation, providing features that make expectations visible, allow roles to be flexibly reassigned, and reduce the hidden labor CPs currently perform in mediating between stakeholders.}

\textcolor{black}{\subsubsection{Establishing a “new normal” that maximizes the dyad’s autonomy and wellbeing}
As a dyad begins to navigate the increasing cognitive challenges \textcolor{black}{experienced by the pwMCI}, they have to re-orient their entire relationship toward this new relationship dynamic, with the CP often taking on roles previously played by the pwMCI. For example, we often found that one person had been previously responsible for tasks such as paying the bills or coordinating and completing home repairs, with those responsibilities now shifting to CPs regardless of previous norms \textcolor{black}{[CC01, Section 4.2.1]}. This resulted in CPs having to take on new responsibilities while needing to ensure that the pwMCI is able to stay engaged as much as possible, depending on the severity of their diagnosis. Prior work discusses this by highlighting the importance of understand\textcolor{black}{ing} the situated context of an MCI dyad, and provid\textcolor{black}{ing} support such that the pwMCI feels \textcolor{black}{in control and maintains their sense of autonomy} while the CP maintains personal wellbeing and peace of mind \cite{madjaroff2017narratives}. While we found this to be true across our participants, we also found that CPs struggle to scaffold the involvement of the pwMCI to maximize their independence. It was clear that maximized autonomy for the pwMCI is ideal, however, identifying and incorporating these means of independence proved difficult. Since some elements of the pwMCI’s individual routine were lost post-diagnosis, CPs would find ways to involve themselves as minimally as possible. This took many forms, such as the CP \textcolor{black}{downloading a speech-to-text app on the pwMCI’s phone so they could send text messages despite having difficulty spelling out words [CC05, Section 4.2.1], as well as} the CP placing visual \textcolor{black}{or audio} reminders in different key places of the home so the pwMCI would remember to do something in that place when the CP wasn’t around [CC10 \textcolor{black}{and CC11, Section 4.1.1}].} \textcolor{black}{In some cases, our findings also point to the gendered nature of care coordination among older adults, particularly when a wife’s MCI disrupted household routines she had previously managed. In these situations, responsibilities that had long been sustained through invisible, gendered labor became explicit, often requiring adult children to step in and assume coordination tasks on behalf of both parents [CC02\textcolor{black}{, Section 4.1.1}]. These observations suggest that the gendered dynamics of care coordination during the transitional MCI phase warrant deeper examination in future work, particularly to understand how roles are renegotiated across family members and over time.}

\textcolor{black}{Another aspect of addressing the new normal within dyads that happens post-diagnosis is the reality that a pwMCI’s support needs, and hence the nature of coordination and orchestration, will \textcolor{black}{continue to} change due to the progressive nature of MCI. The approaches that CPs use to maximize independence and safety evolve with the changing abilities of the diagnosed individual. As a result, CPs must constantly adapt their role in light of such continuous shifts. These kinds of adaptive measures are seen in prior work when a caregiver is tasked with providing longitudinal informal care for their loved one with a chronic and progressive diagnosis \cite{zarit1986subjective, grunfeld2004family}. In these instances, the caregivers find it increasingly difficult to navigate the process of providing appropriate care as they balance the needs of their loved one with the impact of the progressive diagnosis on their own wellbeing. The CPs we interviewed expressed these same thoughts and feelings, reflecting on the emotional challenges that come about as they progress through the experience of longitudinal care. \textcolor{black}{This change was often perceived as feeling like the loss of a former self or former life, and framing the post-diagnosis journey through this grief-like lens helped CPs} better manage the mental toll of navigating coordination activities while being acutely aware of the pwMCI’s decline. These emotional experiences are not peripheral to coordination, but directly shape how CPs plan, delegate, and sustain orchestration over time. Feelings of grief, fatigue, and uncertainty can limit the bandwidth CPs have to experiment with new tools, involve others in the network, or continually adapt routines as needs shift. This underscores the importance of designing future support solutions that meet the day-to-day needs of pwMCI and also ease the orchestration burden on CPs, by reducing the scaffolding required, making adaptations more seamless, and supporting the emotional as well as logistical dimensions of longitudinal care.}

\subsection{Design Opportunities for Future Technology \textcolor{black}{for Orchestrating Informal Care}}

Our findings about the evolving nature of care coordination revealed that the orchestration work performed by CPs is rarely \textcolor{black}{adequately} supported by available technologies. \textcolor{black}{Every} dyad described gaps between their coordination practices and the tools they relied on, whether digital or analog. These gaps, such as the inability to scaffold coordination tasks without first planning every aspect of that task ahead of time, exposed a more fundamental misalignment between how current technologies are designed and how coordination actually unfolds in MCI care networks. Instead of facilitating effective orchestration, many tools left CPs \textcolor{black}{needing} to perform extra planning, re-purposing, or emotional labor. \textcolor{black}{These lived experiences expose not only the shortcomings of current solutions to meet the needs of MCI dyads, but point to opportunities for future technologies to be designed with the situated context of informal care for pwMCI in mind. At-home solutions that adapt to an ever-changing ecosystem of care which revolves around central orchestration by a single, informal care coordinator would significantly ease the challenges faced by CPs as they coordinate day-to-day care. Such a context-aware support system is possible if future designers consider the unique challenges faced by these users and how current technolog\textcolor{black}{ies do not} fully support them.}

\textcolor{black}{\subsubsection{Adapting non-specific technology mediums to support specific coordination needs}
One of the various ways that CPs found themselves engaging in extraneous coordination effort was when they had to utilize tools that weren't specifically designed for their context, leading to a need for repurposing general-purpose technologies for highly specific coordination tasks. For example, \textcolor{black}{one} CP described using \textcolor{black}{an} Alexa-enabled device to scaffold shopping routines by dictating grocery lists in \textcolor{black}{a} way that allowed the pwMCI to participate [CC05\textcolor{black}{, Section 4.1.1}]\textcolor{black}{; another CP set up an} iPad that \textcolor{black}{was} permanently plugged in and uncovered to enable predictable FaceTime calls [CC02\textcolor{black}{, Section 4.1.1}]. These practices highlight the creativity with which CPs “rigged” technologies to maintain autonomy while ensuring coordination could still occur. Our insights point to many ways that current technology is used to solve problems that it wasn’t necessarily created to solve. Whether it be using the reminder function on an Alexa-enabled device to allow a CP to remind a pwMCI of a routine chore that can be done at a time \textcolor{black}{when} the CP knows \textcolor{black}{the pwMCI} will be within earshot of the device [CC11\textcolor{black}{, Section 4.1.1}], or \textcolor{black}{using the geolocation functionality on an iWatch to track the location of a pwMCI when they’re out for a walk [CC09, Section 4.1.1]}, we identified several everyday scenarios \textcolor{black}{where} technology is able to support dyads through scaffolding. Prior work has also observed similar strategies in caregiving: families using calendars and messaging apps to coordinate appointments \cite{yamashita2018information, renyi2022uncovering}, voice assistants repurposed for medication adherence \cite{mathur2022collaborative}, or mobile tools for financial management \cite{chan2025insights, dai2025envisioning} and meal preparation \cite{chan2025insights}.} 

\textcolor{black}{Our findings extend these insights across the entire coordination spectrum of MCI, revealing that the most fragile points in daily routines are often the ones that depend on timing, presence, and reassurance: knowing that reminders will be delivered when the pwMCI can hear them, ensuring communication channels are always available, or verifying safety during independent activities like walking. In these scenarios, generic tools provide partial solutions, but only by requiring CPs to anticipate failure modes and scaffold workarounds, adding to their invisible labor. The design opportunity here lies in developing systems that move beyond piecemeal functionality to support orchestration-aware coordination. Rather than forcing CPs to adapt generic devices, future systems should allow for contextual input from CPs, enabling them to specify what task needs support (e.g., a reminder or check-in) and, also, more importantly, \textit{when} and \textit{how} it will be most effective (e.g., when the pwMCI is present in a specific room). Furthermore, \textcolor{black}{our insights emphasize} the need to integrate multi-functionality into a single platform, so that dyads are not forced to cobble together multiple devices and platforms. This underscores the need for technology design that combines features of existing supports, such as voice activation with delayed action-response, videoconferencing, geolocation that is synced to other kinds of devices, and different voice engagement options that users may want to configure for daily use.}

\textcolor{black}{\subsubsection{Designing for Trust and Progressive Adoption of Support for the “New Normal”}
Technology disuse is a common theme amongst both older adults in general and those with cognitive decline for a variety of reasons including stigmas around tech literacy \cite{guisado2019factors}, perceived unreliability of technology solutions \cite{pradhan2020use}, and general unfamiliarity with technical media that can lead to feelings of untrustworthiness \cite{knowles2018older}. These various forms of complexity around the adoption of technical tools can make the barrier to use significant and often insurmountable, especially in situations where a pwMCI will only use a particular technology if consistently prompted, encouraged, and in some cases pushed by their CP. This combination of distrust and lack of clarity around proper use and usefulness led to widespread lack of interaction with technology amongst pwMCI that we interviewed, and often a lack of desire in CPs to engage new technologies given concerns that the costs would outweigh the benefits.} \textcolor{black}{This shared experience highlights the need for technologies that balance familiarity, ease of use, and value. Such technologies should encourage adoption early in the care coordination process and remain usable as needs evolve over time.}

\textcolor{black}{Building on these insights, our findings suggest several design directions. \textit{First}, technologies should offer progressive disclosure of functionality, allowing dyads to begin with simple, familiar features (e.g., basic reminders or shared notes) and gradually scale to more complex coordination supports as comfort grows. \textit{Second}, systems could incorporate redundant, multi-modal prompts, such as combining auditory cues with physical artifacts like calendars or notes, to ease adoption and preserve existing routines. \textit{Third}, to support orchestration beyond the dyad, tools should provide configurable visibility controls, so CPs can selectively share updates with family or formal providers without overwhelming or alienating the pwMCI. This selective sharing of information among caregivers has also shown to positively affect caregiving outcomes for older adults \cite{crotty2015information}.} 

\textcolor{black}{\subsubsection{From Manual Workarounds to Hybrid Care Coordination Tools}
A notable finding from our interviews was the continued reliance on analog and manual methods, such as whiteboards, paper calendars, and notebooks, to manage care coordination. Participants described these practices as both more efficient and more trustworthy than digital tools, particularly when the technologies introduced confusion, complexity, or excessive time costs. Prior work has documented older adults’ preference for manual methods when digital tools fail to align with everyday practices \cite{dai2025envisioning, chan2025insights}, and how caregivers often “patch together” manual and digital artifacts to create workable systems of support \cite{renyi2022uncovering, yamashita2019associations}. Our findings extend these insights and show how dyads often abandoned or bypassed digital systems not just due to usability barriers, as identified in prior works \cite{knowles2018wisdom, lee2024effects}, but also because analog methods offered them a sense of continuity and predictability that supported their goals of living with imminent cognitive decline.} 

\textcolor{black}{Importantly, participants did not reject digital technologies outright. Instead, they fluidly moved between analog and digital modalities depending on task, context and cognitive load. For example, paper calendars or whiteboards served as stable and accessible anchors for daily routines, while digital tools were selectively layered on top to provide reminders, remote access or backup tracking. This insight also speaks to prior work in medication management for older adults, where Mathur et al. demonstrate the value of supporting alternative and multiple interaction \textcolor{black}{modalities} at the same time (such as combining voice, visual and physical artifacts together) to align with older adults’ existing practices rather than replacing them \cite{mathur2022collaborative}. Our findings suggest that similar hybrid approaches are critical even beyond medication tasks, spanning scheduling, communication, safety monitoring and household coordination.} \textcolor{black}{\textcolor{black}{Additionally, \textcolor{black}{this} reliance on manual practices highlights a fundamental gap in current systems: most technologies assume that increased automation will reduce coordination effort, yet often overlook the orchestration role of CPs.} In practice, fully automated systems often add to CPs’ mental burden, requiring additional oversight and troubleshooting, whereas analog approaches allow CPs to offload vigilance onto the routines they trust. These experiences reveal that CPs and pwMCI do not simply prefer paper calendars or whiteboards out of habit, they turn to them because existing technologies fail to align with the realities of orchestration work, introducing more labor. By surfacing these dynamics, our analysis extends prior accounts of caregiving orchestration to show how analog practices remain essential in sustaining daily life with MCI.} 

\textcolor{black}{A key implication of our findings is that the feasibility and usefulness of coordination technologies for pwMCI are inherently shaped by the transitional and unpredictable nature of cognitive decline. As individuals move through different stages of MCI, their ability to directly engage with technologies (and the types of interaction modalities that remain accessible) can change substantially. Rather than viewing this as a limitation of the proposed design directions, our findings suggest that effective systems must be explicitly designed for transition. In practice, this means supporting a gradual shift in agency from the pwMCI to the CP, without requiring abrupt abandonment or replacement of tools as cognitive abilities change. For example, technologies that initially foreground pwMCI participation through familiar, low-effort interactions (e.g., voice prompts, shared visual artifacts) may later function primarily as background infrastructure for CP-led orchestration, such as safety monitoring, scheduling or coordination with others. In this sense, the value of these systems \textcolor{black}{lies in} their ability to adapt alongside the dyad, while preserving continuity, reducing reconfiguration costs and supporting care partners as coordination increasingly becomes anticipatory rather than collaborative.}

\textcolor{black}{Our work also challenges a common notion about older adults in technological research, one that characterizes them as \textit{passive recipients of technology}. We argue that instead of looking at older adults’ creative means of adopting and negotiating with digital tools as purely technological resistance, designers of such systems should reframe these actions as meaningful design signals and opportunities. Building on work that \textcolor{black}{delineates} the importance of scaffolding autonomy in dementia and MCI contexts \cite{mathur2022collaborative, zubatiy2021empowering, schurgin2021isolation}, we \textcolor{black}{emphasize that} CPs need socio-technical systems that integrate with, rather than attempt to replace, the analog scaffolds that structure their daily lives. Future research should therefore explore hybrid designs that bridge analog and digital practices. For example, systems might digitize handwritten notes or automatically synchronize wall calendars with shared digital reminders \cite{beneteau2019communication}, ensuring continuity while reducing redundant effort. Such designs would preserve familiarity and autonomy for pwMCI and also relieve CPs of the constant invisible labor required to maintain coordination.}

\section{Limitations}
While this work takes important steps toward a better understanding of the \textcolor{black}{informal care coordination} needs of people diagnosed with MCI and their CPs, it exposes the need to further explore more nuanced aspects of these day-to-day realities. For example, we found strong evidence for the many unique ways that CPs are able to orchestrate daily needs for a pwMCI \textcolor{black}{via their own efforts and the assistance of their larger care network}. However, further exploration is needed to quantify the ways in which CPs both successfully and unsuccessfully leverage technology to \textcolor{black}{plan for and manage} coordination tasks \textcolor{black}{in an emotionally aware way\textcolor{black}{,} }beyond the high-level themes we identified. Our approach also limited the input we were able to receive directly from pwMCI, as CPs would either speak to us alone on behalf of a pwMCI, or a pwMCI would not provide much commentary when asked questions about their routines, activities, experiences, and opinions. A mixed-methods exploration into their day-to-day life may resolve this limitation by giving pwMCI greater opportunity to provide input\textcolor{black}{, or by allowing the dyad an opportunity to capture their experiences in the moment through a variety of available ethnographic approaches. Such an approach would yield a more holistic representation of the inherent complexity and nuance of day-to-day supports, routines, and care coordination practices.} \textcolor{black}{Finally, the implications of this work must be considered within the context of our participant population, which includes dyads consisting of an individual with MCI and a primary care partner \textcolor{black}{who is} either a spouse\textcolor{black}{,} an adult child\textcolor{black}{, or other family member}. Many individuals with MCI live alone or do not have a single dedicated care partner who can support their day-to-day needs on a regular basis. Therefore, our findings may not readily generalize to these isolated individuals who likely have more fragmented and de-centralized care networks, presenting a promising potential future direction for work on informal care coordination in MCI.}

\section{Conclusion}

The prevalence of cognitive decline among older adults and the resulting \textcolor{black}{need for informal care delivered by family members and other loved ones} poses a unique set of challenges. \textcolor{black}{A key focus of informal caregiving for MCI is to maximize independence and quality of life through the provision of daily supports and emotionally attuned care. Our} work provides key insights needed to \textcolor{black}{design technological} solutions \textcolor{black}{to support care partners involved in orchestrating these supports, leveraging interview data to unpack the current means by which informal coordination takes place.} These results offer opportunities for the design of future support systems that center the needs of informal care networks. In doing so, this study lays the groundwork for future explorations into informal MCI care \textcolor{black}{coordination} and encourages a human-centric approach when creating solutions for the growing aging population.

\begin{acks}
    We would like to thank all the participants involved in this study for their time and for sharing their experiences. This material is based upon work supported by the National AI Research Institutes program, supported by the National Science Foundation (NSF) in partnership under Award No. 2112633. Any opinions, findings, and conclusions or recommendations expressed in this material are those of the author(s) and do not necessarily reflect the views of the NSF.  
\end{acks}

\bibliographystyle{ACM-Reference-Format}
\bibliography{main}

@article{johnson2020roles,
  title={Roles in the discussion: an analysis of social support in an online forum for people with dementia},
  author={Johnson, Jazette and Black, Rebecca W and Hayes, Gillian R},
  journal={Proceedings of the ACM on Human-Computer Interaction},
  volume={4},
  number={CSCW2},
  pages={1--30},
  year={2020},
  publisher={ACM New York, NY, USA}
}

@article{mathur2025sometimes,
  title={" Sometimes You Need Facts, and Sometimes a Hug": Understanding Older Adults' Preferences for Explanations in LLM-Based Conversational AI Systems},
  author={Mathur, Niharika and Zubatiy, Tamara and Rozga, Agata and Forlizzi, Jodi and Mynatt, Elizabeth},
  journal={arXiv preprint arXiv:2510.06697},
  year={2025}
}

@article{busse2006mild,
  title={Mild cognitive impairment: long-term course of four clinical subtypes},
  author={Busse, Anja and Hensel, Anke and Guhne, Uta and Angermeyer, Matthias C and Riedel-Heller, Steffi G},
  journal={Neurology},
  volume={67},
  number={12},
  pages={2176--2185},
  year={2006},
  publisher={Lippincott Williams \& Wilkins}
}

@article{fischer2007conversion,
  title={Conversion from subtypes of mild cognitive impairment to Alzheimer dementia},
  author={Fischer, Peter and Jungwirth, Susanne and Zehetmayer, Sonja and Weissgram, S and Hoenigschnabl, Sebastian and Gelpi, Ellen and Krampla, Wolfgang and Tragl, Karl Heinz},
  journal={Neurology},
  volume={68},
  number={4},
  pages={288--291},
  year={2007},
  publisher={Lippincott Williams \& Wilkins}
}

@article{pashmdarfard2020assessment,
  title={Assessment tools to evaluate Activities of Daily Living (ADL) and Instrumental Activities of Daily Living (IADL) in older adults: A systematic review},
  author={Pashmdarfard, Marzieh and Azad, Akram},
  journal={Medical journal of the Islamic Republic of Iran},
  volume={34},
  pages={33},
  year={2020}
}

@article{kim2025evaluation,
  title={Evaluation of activities of daily living: current insights and future horizons},
  author={Kim, Jin-Ho and Lee, Seok Bum},
  journal={Annals of Geriatric Medicine and Research},
  volume={29},
  number={2},
  pages={143},
  year={2025}
}

@article{zhou2025adhera,
  title={Adhera: A Human-Centered Health Informatics Solution for Reducing Informal Caregiver Burden through Improved Medication Adherence},
  author={Zhou, Zhiyin},
  journal={arXiv preprint arXiv:2512.03878},
  year={2025}
}

@article{knodel2010role,
  title={The role of parents and family members in ART treatment adherence: Evidence from Thailand},
  author={Knodel, John and Kespichayawattana, Jiraporn and Saengtienchai, Chanpen and Wiwatwanich, Suvinee},
  journal={Research on Aging},
  volume={32},
  number={1},
  pages={19--39},
  year={2010},
  publisher={SAGE Publications Sage CA: Los Angeles, CA}
}

@article{dawber2019comparison,
  title={Comparison of informal caregiver and named nurse assessment of symptoms in elderly patients dying in hospital using the palliative outcome scale},
  author={Dawber, Rebecca and Armour, Kathy and Ferry, Peter and Mukherjee, Bhaskar and Carter, Christopher and Meystre, Chantal},
  journal={BMJ supportive \& palliative care},
  volume={9},
  number={2},
  pages={175--182},
  year={2019},
  publisher={British Medical Journal Publishing Group}
}

@article{ullgren2018family,
  title={How family caregivers of cancer patients manage symptoms at home: a systematic review},
  author={Ullgren, Helena and Tsitsi, Theologia and Papastavrou, Evridiki and Charalambous, Andreas},
  journal={International journal of nursing studies},
  volume={85},
  pages={68--79},
  year={2018},
  publisher={Elsevier}
}

@article{zarit2008behavioral,
  title={Behavioral and psychosocial interventions for family caregivers},
  author={Zarit, Steven and Femia, Elia},
  journal={Journal of Social Work Education},
  volume={44},
  number={sup3},
  pages={49--57},
  year={2008},
  publisher={Taylor \& Francis}
}

@article{ducharme2011learning,
  title={“Learning to become a family caregiver” efficacy of an intervention program for caregivers following diagnosis of dementia in a relative},
  author={Ducharme, Francine C and L{\'e}vesque, Louise L and Lachance, Lise M and Kergoat, Marie-Jeanne and Legault, Alain J and Beaudet, Line M and Zarit, Steven H},
  journal={The Gerontologist},
  volume={51},
  number={4},
  pages={484--494},
  year={2011},
  publisher={Oxford University Press}
}

@inproceedings{chen2013caring,
  title={Caring for caregivers: designing for integrality},
  author={Chen, Yunan and Ngo, Victor and Park, Sun Young},
  booktitle={Proceedings of the 2013 conference on Computer supported cooperative work},
  pages={91--102},
  year={2013}
}

@article{nikkhah2024family,
  title={Family Resilience in Care Coordination Technologies: Designing for Families as Adaptive Systems},
  author={Nikkhah, Sarah and Rode, Akash Uday and Kulkarni, Neha Keshav and Mittal, Priyanjali and Mueller, Emily L and Miller, Andrew D},
  journal={Proceedings of the ACM on Human-Computer Interaction},
  volume={8},
  number={CSCW2},
  pages={1--28},
  year={2024},
  publisher={ACM New York, NY, USA}
}

@inproceedings{currin2019give,
  title={Give me a break: design for communication among family caregivers and respite caregivers},
  author={Currin, Flannery and Razo, Gustavo and Min, Aehong},
  booktitle={Extended Abstracts of the 2019 CHI Conference on Human Factors in Computing Systems},
  pages={1--6},
  year={2019}
}

@inproceedings{miller2016partners,
  title={Partners in care: design considerations for caregivers and patients during a hospital stay},
  author={Miller, Andrew D and Mishra, Sonali R and Kendall, Logan and Haldar, Shefali and Pollack, Ari H and Pratt, Wanda},
  booktitle={Proceedings of the 19th ACM Conference on Computer-Supported Cooperative Work \& Social Computing},
  pages={756--769},
  year={2016}
}

@inproceedings{foong2020you,
  title={" You Cannot Offer Such a Suggestion" Designing for Family Caregiver Input in Home Care Systems},
  author={Foong, Pin Sym and Lim, Charis Anne and Wong, Joshua and Lim, Chang Siang and Perrault, Simon Tangi and Koh, Gerald CH},
  booktitle={Proceedings of the 2020 CHI Conference on Human Factors in Computing Systems},
  pages={1--13},
  year={2020}
}

@article{bhat2023we,
  title={" We are half-doctors": family caregivers as boundary actors in chronic disease management},
  author={Bhat, Karthik S and Hall, Amanda K and Kuo, Tiffany and Kumar, Neha},
  journal={Proceedings of the ACM on human-computer interaction},
  volume={7},
  number={CSCW1},
  pages={1--29},
  year={2023},
  publisher={ACM New York, NY, USA}
}

@article{grunfeld2004family,
  title={Family caregiver burden: results of a longitudinal study of breast cancer patients and their principal caregivers},
  author={Grunfeld, Eva and Coyle, Doug and Whelan, Timothy and Clinch, Jennifer and Reyno, Leonard and Earle, Craig C and Willan, Andrew and Viola, Raymond and Coristine, Marjorie and Janz, Teresa and others},
  journal={Cmaj},
  volume={170},
  number={12},
  pages={1795--1801},
  year={2004},
  publisher={CMAJ}
}

@article{guisado2019factors,
  title={Factors influencing the adoption of smart health technologies for people with dementia and their informal caregivers: scoping review and design framework},
  author={Guisado-Fern{\'a}ndez, Estefan{\'\i}a and Giunti, Guido and Mackey, Laura M and Blake, Catherine and Caulfield, Brian Michael},
  journal={JMIR aging},
  volume={2},
  number={1},
  pages={e12192},
  year={2019},
  publisher={JMIR Publications Inc., Toronto, Canada}
}

@article{pradhan2020use,
  title={Use of intelligent voice assistants by older adults with low technology use},
  author={Pradhan, Alisha and Lazar, Amanda and Findlater, Leah},
  journal={ACM Transactions on Computer-Human Interaction (TOCHI)},
  volume={27},
  number={4},
  pages={1--27},
  year={2020},
  publisher={ACM New York, NY, USA}
}

@inproceedings{chan2025insights,
  title={Insights from Designing Context-Aware Meal Preparation Assistance for Older Adults with Mild Cognitive Impairment (MCI) and Their Care Partners},
  author={Chan, Szeyi and Li, Jiachen and Ao, Siman and Wang, Yufei and Bilau, Ibrahim and Jones, Brian D and Yang, Eunhwa and Mynatt, Elizabeth D and Tan, Xiang Zhi},
  booktitle={Proceedings of the 2025 ACM Designing Interactive Systems Conference},
  pages={3263--3279},
  year={2025}
}

@inproceedings{dai2025envisioning,
  title={Envisioning Financial Technology Support for Older Adults Through Cognitive and Life Transitions},
  author={Dai, Jiamin and McGrenere, Joanna},
  booktitle={Proceedings of the 2025 CHI Conference on Human Factors in Computing Systems},
  pages={1--24},
  year={2025}
}

@article{zarit1986subjective,
  title={Subjective burden of husbands and wives as caregivers: a longitudinal study},
  author={Zarit, Steven H and Todd, Pamela A and Zarit, Judy M},
  journal={The Gerontologist},
  volume={26},
  number={3},
  pages={260--266},
  year={1986},
  publisher={The Gerontological Society of America}
}

@article{devries2019impact,
  title={The impact of reading groups on engagement and social interaction for older adults with dementia: A literature review},
  author={DeVries, Dawn and Bollin, Angela and Brouwer, Karley and Marion, Alexandra and Nass, Hannah and Pompilius, Amanda},
  journal={Therapeutic Recreation Journal},
  volume={53},
  number={1},
  pages={53--75},
  year={2019},
  publisher={Sagamore Publishing LLC}
}

@article{crotty2015information,
  title={Information sharing preferences of older patients and their families},
  author={Crotty, Bradley H and Walker, Jan and Dierks, Meghan and Lipsitz, Lewis and O’Brien, Jacqueline and Fischer, Shira and Slack, Warner V and Safran, Charles},
  journal={JAMA internal medicine},
  volume={175},
  number={9},
  pages={1492--1497},
  year={2015},
  publisher={American Medical Association}
}

@article{renyi2022uncovering,
  title={Uncovering the complexity of care networks--towards a taxonomy of collaboration complexity in homecare},
  author={Renyi, Madeleine and Gaugisch, Petra and Hunck, Alexandra and Strunck, Stefan and Kunze, Christophe and Teuteberg, Frank},
  journal={Computer Supported Cooperative Work (CSCW)},
  volume={31},
  number={3},
  pages={517--554},
  year={2022},
  publisher={Springer}
}

@article{knowles2018wisdom,
  title={The wisdom of older technology (non) users},
  author={Knowles, Bran and Hanson, Vicki L},
  journal={Communications of the ACM},
  volume={61},
  number={3},
  pages={72--77},
  year={2018},
  publisher={ACM New York, NY, USA}
}

@article{lee2024effects,
  title={Effects of Mobile Health Applications in Older Adults with Dementia or Mild Cognitive Impairment: A Systematic Review and Meta-Analysis},
  author={Lee, Minjae and Park, Jisung and Lee, Seunghyeon},
  journal={Korean Journal of Adult Nursing},
  volume={36},
  number={2},
  pages={112--125},
  year={2024}
}

@article{yamashita2019associations,
  title={Associations between motivation to learn, basic skills, and adult education and training participation among older adults in the USA},
  author={Yamashita, Takashi and Cummins, Phyllis A and Millar, Roberto J and Sahoo, Shalini and Smith, Thomas J},
  journal={International Journal of Lifelong Education},
  volume={38},
  number={5},
  pages={538--552},
  year={2019},
  publisher={Taylor \& Francis}
}

@inproceedings{beneteau2019communication,
  title={Communication breakdowns between families and Alexa},
  author={Beneteau, Erin and Richards, Olivia K and Zhang, Mingrui and Kientz, Julie A and Yip, Jason and Hiniker, Alexis},
  booktitle={Proceedings of the 2019 CHI conference on human factors in computing systems},
  pages={1--13},
  year={2019}
}

@article{mank2023determinants,
  title={Determinants of informal care time, distress, depression, and quality of life in care partners along the trajectory of Alzheimer's disease},
  author={Mank, Arenda and van Maurik, Ingrid S and Rijnhart, Judith JM and Rhodius-Meester, Hanneke FM and Visser, Leonie NC and Lemstra, Afina W and Sikkes, Sietske AM and Teunissen, Charlotte E and van Giessen, Elsmarieke M and Berkhof, Johannes and others},
  journal={Alzheimer's \& Dementia: Diagnosis, Assessment \& Disease Monitoring},
  volume={15},
  number={2},
  pages={e12418},
  year={2023},
  publisher={Wiley Online Library}
}

@article{sherman2017efficacy,
  title={The efficacy of cognitive intervention in mild cognitive impairment (MCI): a meta-analysis of outcomes on neuropsychological measures},
  author={Sherman, Dale S and Mauser, Justin and Nuno, Miriam and Sherzai, Dean},
  journal={Neuropsychology review},
  volume={27},
  pages={440--484},
  year={2017},
  publisher={Springer}
}

@article{albert2011diagnosis,
  title={The diagnosis of mild cognitive impairment due to Alzheimer's disease: recommendations from the National Institute on Aging-Alzheimer's Association workgroups on diagnostic guidelines for Alzheimer's disease},
  author={Albert, Marilyn S and DeKosky, Steven T and Dickson, Dennis and Dubois, Bruno and Feldman, Howard H and Fox, Nick C and Gamst, Anthony and Holtzman, David M and Jagust, William J and Petersen, Ronald C and others},
  journal={Alzheimer's \& dementia},
  volume={7},
  number={3},
  pages={270--279},
  year={2011},
  publisher={Wiley Online Library}
}

@article{chandler2019comparative,
  title={Comparative effectiveness of behavioral interventions on quality of life for older adults with mild cognitive impairment: a randomized clinical trial},
  author={Chandler, Melanie J and Locke, Dona E and Crook, Julia E and Fields, Julie A and Ball, Colleen T and Phatak, Vaishali S and Dean, Pamela M and Morris, Miranda and Smith, Glenn E},
  journal={JAMA network open},
  volume={2},
  number={5},
  pages={e193016--e193016},
  year={2019},
  publisher={American Medical Association}
}

@article{knowles2018older,
  title={Older adults’ deployment of ‘distrust’},
  author={Knowles, Bran and Hanson, Vicki L},
  journal={ACM Transactions on Computer-Human Interaction (TOCHI)},
  volume={25},
  number={4},
  pages={1--25},
  year={2018},
  publisher={ACM New York, NY, USA}
}

@inproceedings{gutierrez2017takes,
  title={It takes at least two to tango: understanding the cooperative nature of elderly caregiving in Latin America},
  author={Gutierrez, Francisco J and Ochoa, Sergio F},
  booktitle={Proceedings of the 2017 ACM Conference on computer supported cooperative work and social computing},
  pages={1618--1630},
  year={2017}
}

@article{consolvo2004technology,
  title={Technology for care networks of elders},
  author={Consolvo, Sunny and Roessler, Peter and Shelton, Brett E and LaMarca, Anthony and Schilit, Bill and Bly, Sara},
  journal={IEEE pervasive computing},
  volume={3},
  number={2},
  pages={22--29},
  year={2004},
  publisher={IEEE}
}

@inproceedings{yamashita2018information,
  title={How information sharing about care recipients by family caregivers impacts family communication},
  author={Yamashita, Naomi and Kuzuoka, Hideaki and Kudo, Takashi and Hirata, Keiji and Aramaki, Eiji and Hattori, Kazuki},
  booktitle={Proceedings of the 2018 chi conference on human factors in computing systems},
  pages={1--13},
  year={2018}
}

@article{tang2018awareness,
  title={Awareness and handoffs in home care: coordination among informal caregivers},
  author={Tang, Charlotte and Chen, Yunan and Cheng, Karen and Ngo, Victor and Mattison, John E},
  journal={Behaviour \& Information Technology},
  volume={37},
  number={1},
  pages={66--86},
  year={2018},
  publisher={Taylor \& Francis}
}

@article{dean2012living,
  title={Living with mild cognitive impairment: the patient's and carer's experience},
  author={Dean, Katherine and Wilcock, Gordon},
  journal={International psychogeriatrics},
  volume={24},
  number={6},
  pages={871--881},
  year={2012},
  publisher={Cambridge University Press}
}

@article{austrom2009long,
  title={Long term caregiving: helping families of persons with mild cognitive impairment cope},
  author={Austrom, Mary G and Lu, Yvonne},
  journal={Current Alzheimer Research},
  volume={6},
  number={4},
  pages={392--398},
  year={2009},
  publisher={Bentham Science Publishers direct}
}

@article{alm2002designing,
  title={Designing an interface usable by people with dementia},
  author={Alm, Norman and Dye, Richard and Gowans, Gary and Campbell, Jim and Astell, Arlene and Ellis, Maggie},
  journal={ACM SIGCAPH Computers and the Physically Handicapped},
  number={73-74},
  pages={156--157},
  year={2002},
  publisher={ACM New York, NY, USA}
}

@inproceedings{kuwahara2006networked,
  title={Networked reminiscence therapy for individuals with dementia by using photo and video sharing},
  author={Kuwahara, Noriaki and Abe, Shinji and Yasuda, Kiyoshi and Kuwabara, Kazuhiro},
  booktitle={Proceedings of the 8th international ACM SIGACCESS conference on Computers and accessibility},
  pages={125--132},
  year={2006}
}

@inproceedings{zaccarelli2013computer,
  title={Computer-based cognitive intervention for dementia: sociable: motivating platform for elderly networking, mental reinforcement and social interaction},
  author={Zaccarelli, Chiara and Cirillo, Giulio and Passuti, Simone and Annicchiarico, Roberta and Barban, Francesco},
  booktitle={Proceedings of the 7th international conference on pervasive computing technologies for healthcare},
  pages={430--435},
  year={2013}
}

@article{adams2006transition,
  title={The transition to caregiving: the experience of family members embarking on the dementia caregiving career},
  author={Adams, Kathryn Betts},
  journal={Journal of gerontological social work},
  volume={47},
  number={3-4},
  pages={3--29},
  year={2006},
  publisher={Taylor \& Francis}
}

@article{corbin1985managing,
  title={Managing chronic illness at home: three lines of work},
  author={Corbin, Juliet and Strauss, Anselm},
  journal={Qualitative sociology},
  volume={8},
  number={3},
  pages={224--247},
  year={1985},
  publisher={Springer}
}

@article{yang2022magic,
  title={Magic Brush: An AI-based Service for Dementia Prevention focused on Intrinsic Motivation.},
  author={Yang, Migyeong and Lee, Kyungha and Kim, Eunji and Song, Yeosol and Lee, Sewang and Kang, Jiwon and Han, Jinyoung and Song, Hayeon and Kim, Taeeun},
  journal={Proc. ACM Hum. Comput. Interact.},
  volume={6},
  number={CSCW2},
  pages={1--21},
  year={2022}
}

@article{lydon2022integrative,
  title={An integrative framework to guide social engagement interventions and technology design for persons with mild cognitive impairment},
  author={Lydon, Elizabeth A and Nguyen, Lydia T and Nie, Qiong and Rogers, Wendy A and Mudar, Raksha A},
  journal={Frontiers in public health},
  volume={9},
  pages={750340},
  year={2022},
  publisher={Frontiers Media SA}
}

@article{lazar2018negotiating,
  title={Negotiating relation work with telehealth home care companionship technologies that support aging in place},
  author={Lazar, Amanda and Thompson, Hilaire J and Lin, Shih-Yin and Demiris, George},
  journal={Proceedings of the ACM on Human-Computer Interaction},
  volume={2},
  number={CSCW},
  pages={1--19},
  year={2018},
  publisher={ACM New York, NY, USA}
}

@inproceedings{zubatiy2021empowering,
  title={Empowering dyads of older adults with mild cognitive impairment and their care partners using conversational agents},
  author={Zubatiy, Tamara and Vickers, Kayci L and Mathur, Niharika and Mynatt, Elizabeth D},
  booktitle={Proceedings of the 2021 CHI conference on human factors in computing systems},
  pages={1--15},
  year={2021}
}

@article{zubatiy2023don,
  title={" I don't know how to help with that"-Learning from Limitations of Modern Conversational Agent Systems in Caregiving Networks},
  author={Zubatiy, Tamara and Mathur, Niharika and Heck, Larry and Vickers, Kayci L and Rozga, Agata and Mynatt, Elizabeth D},
  journal={Proceedings of the ACM on Human-Computer Interaction},
  volume={7},
  number={CSCW2},
  pages={1--28},
  year={2023},
  publisher={ACM New York, NY, USA}
}

@inproceedings{mathur2022collaborative,
  title={A collaborative approach to support medication management in older adults with mild cognitive impairment using conversational assistants (CAs)},
  author={Mathur, Niharika and Dhodapkar, Kunal and Zubatiy, Tamara and Li, Jiachen and Jones, Brian and Mynatt, Elizabeth},
  booktitle={Proceedings of the 24th International ACM SIGACCESS Conference on Computers and Accessibility},
  pages={1--14},
  year={2022}
}

@inproceedings{madjaroff2017narratives,
  title={Narratives of older adults with mild cognitive impairment and their caregivers},
  author={Madjaroff, Galina and Mentis, Helena},
  booktitle={Proceedings of the 19th international ACM SIGACCESS conference on computers and accessibility},
  pages={140--149},
  year={2017}
}

@inproceedings{schurgin2021isolation,
  title={Isolation in Coordination: Challenges of Caregivers in the USA},
  author={Schurgin, Mark and Schlager, Mark and Vardoulakis, Laura and Pina, Laura R and Wilcox, Lauren},
  booktitle={Proceedings of the 2021 CHI Conference on Human Factors in Computing Systems},
  pages={1--14},
  year={2021}
}

@article{smriti2024emotion,
  title={Emotion work in caregiving: the role of technology to support informal caregivers of persons living with dementia},
  author={Smriti, Diva and Wang, Lu and Huh-Yoo, Jina},
  journal={Proceedings of the ACM on human-computer interaction},
  volume={8},
  number={CSCW1},
  pages={1--34},
  year={2024},
  publisher={ACM New York, NY, USA}
}

@article{hedman2013patterns,
  title={Patterns of functioning in older adults with mild cognitive impairment: a two-year study focusing on everyday technology use},
  author={Hedman, Annicka and Nyg{\aa}rd, Louise and Almkvist, Ove and Kottorp, Anders},
  journal={Aging \& mental health},
  volume={17},
  number={6},
  pages={679--688},
  year={2013},
  publisher={Taylor \& Francis}
}

@article{lara2019functional,
  title={The functional ability of MCI and Alzheimer’s patients predicts caregiver burden},
  author={Lara-Ruiz, Jose and Kauzor, Kaitlyn and Gonzalez, Katie and Nakhla, Marina Z and Banuelos, Dayana and Woo, Ellen and Apostolova, Liana G and Razani, Jill},
  journal={GeroPsych},
  year={2019},
  publisher={Hogrefe AG}
}

@article{anderson2019state,
  title={State of the science on mild cognitive impairment (MCI)},
  author={Anderson, Nicole D},
  journal={CNS spectrums},
  volume={24},
  number={1},
  pages={78--87},
  year={2019},
  publisher={Cambridge University Press}
}

@article{busse2006progression,
  title={Progression of mild cognitive impairment to dementia: a challenge to current thinking},
  author={Busse, Anja and Angermeyer, Matthias C and Riedel-Heller, Steffi G},
  journal={The British Journal of Psychiatry},
  volume={189},
  number={5},
  pages={399--404},
  year={2006},
  publisher={Cambridge University Press}
}

@article{liang2019optimal,
  title={The optimal treatment for improving cognitive function in elder people with mild cognitive impairment incorporating Bayesian network meta-analysis and systematic review},
  author={Liang, Jing-hong and Shen, Wan-ting and Li, Jia-yu and Qu, Xin-yuan and Li, Jing and Jia, Rui-xia and Wang, Ying-quan and Wang, Shan and Wu, Rong-kun and Zhang, Hong-bo and others},
  journal={Ageing Research Reviews},
  volume={51},
  pages={85--96},
  year={2019},
  publisher={Elsevier}
}

@article{petersen2014mild,
  title={Mild cognitive impairment: a concept in evolution},
  author={Petersen, Ronald C and Caracciolo, Barbara and Brayne, Carol and Gauthier, Serge and Jelic, Vesna and Fratiglioni, Laura},
  journal={Journal of internal medicine},
  volume={275},
  number={3},
  pages={214--228},
  year={2014},
  publisher={Wiley Online Library}
}

@article{petersen1997aging,
  title={Aging, memory, and mild cognitive impairment},
  author={Petersen, Ronald C and Smith, Glenn E and Waring, Stephen C and Ivnik, Robert J and Kokmen, Emre and Tangelos, Eric G},
  journal={International psychogeriatrics},
  volume={9},
  number={S1},
  pages={65--69},
  year={1997},
  publisher={Cambridge University Press}
}

@article{song2023evidence,
  title={Evidence from a meta-analysis and systematic review reveals the global prevalence of mild cognitive impairment},
  author={Song, Wen-xin and Wu, Wei-wei and Zhao, Yuan-yuan and Xu, Hai-lun and Chen, Guan-cheng and Jin, Shan-yu and Chen, Jie and Xian, Shao-xiang and Liang, Jing-hong},
  journal={Frontiers in Aging Neuroscience},
  volume={15},
  pages={1227112},
  year={2023},
  publisher={Frontiers Media SA}
}

@article{wang2022management,
  title={The management of dementia worldwide: a review on policy practices, clinical guidelines, end-of-life care, and challenge along with aging population},
  author={Wang, Changying and Song, Peipei and Niu, Yuhong},
  journal={Bioscience trends},
  volume={16},
  number={2},
  pages={119--129},
  year={2022},
  publisher={International Research and Cooperation Association for Bio \& Socio-Sciences~…}
}

@article{winblad2016defeating,
  title={Defeating Alzheimer's disease and other dementias: a priority for European science and society},
  author={Winblad, Bengt and Amouyel, Philippe and Andrieu, Sandrine and Ballard, Clive and Brayne, Carol and Brodaty, Henry and Cedazo-Minguez, Angel and Dubois, Bruno and Edvardsson, David and Feldman, Howard and others},
  journal={The Lancet Neurology},
  volume={15},
  number={5},
  pages={455--532},
  year={2016},
  publisher={Elsevier}
}


\end{document}